\documentclass[preprint,10pt,twocolumn,superscriptaddress,notitlepage,longbibliography]{revtex4-1}

\usepackage{array}
\usepackage{microtype}
\usepackage{natbib}

\usepackage{graphicx}
\usepackage{amssymb}
\usepackage{multirow}
\usepackage{color}
\usepackage{amsmath}
\usepackage{threeparttable}
\usepackage[nolist,nohyperlinks]{acronym}
\usepackage{hyperref}
\usepackage[nameinlink,capitalize]{cleveref} 
\usepackage{placeins}

\graphicspath{./figures/}

\newcommand{\ue}{\text{e}}
\newcommand{\ud}{\text{d}}

\newcommand{\ie}{\textit{i.e.}}
\newcommand{\eg}{\textit{e.g.}}
\newcommand{\etal}{\textit{et al.}}

\begin{document}

\title{Relaxation Dynamics of Entangled Linear Polymer Melts via Molecular Dynamics Simulations}

\author{Alireza F. Behbahani}
\email{aforooza@uni-mainz.de}
\affiliation{Institut f\"{u}r Physik, Johannes Gutenberg-Universit\"{a}t Mainz, Staudingerweg 7, D-55099 Mainz, Germany}

\author{Friederike Schmid}
\email{friederike.schmid@uni-mainz.de}
\affiliation{Institut f\"{u}r Physik, Johannes Gutenberg-Universit\"{a}t Mainz, Staudingerweg 7, D-55099 Mainz, Germany}

\begin{abstract}
We present an extensive analysis of the relaxation dynamics of
entangled linear polymer melts via long-time molecular dynamics
simulations of a generic bead-spring model. We study the
mean-squared displacements, the autocorrelation
function of the end-to-end vector, $P(t)$, the single-chain
dynamic structure factor, $S(q,t)$, and the linear viscoelastic
properties, especially the shear stress relaxation modulus,
$G(t)$.  
The simulation data are compared with the theoretically expected
scaling laws for different time regimes of entangled melts, and with
analytical expressions that account for different relaxation
mechanisms in the tube model, 
namely, reptation, contour length fluctuation (CLF), and
constraint release (CR).
CLF involves a $t^{1/4}$ scaling regime  in the time-dependence of 
$(1-P(t))$. With increasing chain length, a gradual development
of this scaling regime is observed.  In the absence of CR, the tube
model further predicts that at long times, the chain dynamics
is governed by one central quantity, the ``surviving tube fraction''
$\mu(t)$. As a result, one expects $S(q,t) \propto G(t) \propto P(t)$
in that time regime. We test this prediction by comparing
$S(q,t)$ and $G(t)$ with $P(t)$. For both quantities, 
 proportionality with $P(t)$ is not observed, indicating that CR
has an important effect
on the relaxation of these two quantities. Instead, to a very good
approximation, we find $G(t)\propto P(t)^{2}$ at late times,
which is consistent with the dynamic tube dilation or double
reptation approximations for the CR process.  In addition, we 
calculate non-local mobility functions, which can be used in dynamic
density functional theories for entangled inhomogeneous polymer blends,
and discuss the effect of entanglements on the shape of these
functions.

\end{abstract}
 
 \maketitle
 
\section{Introduction}   
\label{sec:intro}

The simplest basic model of linear polymer dynamics,
the Rouse model, describes
a single Gaussian bead-spring chain that performs unrestricted
Brownian movement in a viscous medium.    
This simple model does not capture
the dynamics of long polymers in a melt because chains cannot
cross each other, and therefore the surrounding chains strongly
restrict the movement of a target chain. To account for this, Edwards, de Gennes, and Doi proposed the so-called tube model, which replaces
the effect of the surrounding chains 
by a confining tube along the contour of the target
chain~\cite{de1971reptation, doi1988theory}.  The tube restricts the
displacement of monomers perpendicular to the local axis of the tube, to a distance called tube diameter,
but the motion of the chain 
along the contour of the tube is not restricted.  
In the basic tube model, the curvilinear diffusive motion
 of the chain along the tube, 
called reptation, 
is assumed to be the (only) mechanism by which the chain 
escapes from the tube and adopts a new conformation~\cite{de1971reptation, doi1988theory,
rubinstein2003polymer}.  However, it is well-known that the pure
reptation mechanism cannot quantitatively explain various features of
entangled polymer melt dynamics, such as the 
scaling exponents of the zero-shear viscosity and the
diffusion coefficient as a function of chain
length\cite{rubinstein2003polymer, mcleish2002tube}. 

To describe polymer melt dynamics more quantitatively, two
additional relaxation mechanisms have been taken into account.  The first mechanism is contour length fluctuation (CLF), which 
takes into account
incoherent chain movements inside the tube (movements other than
reptation, which is the chain's coherent curvilinear motion).  
CLF causes part of the chain to exit the tube
at short times and reduces the effective distance that the chain needs
to travel via reptation
to completely escape from the
tube~\cite{mcleish2002tube,rubinstein2003polymer,doi1996introduction}.
The incorporation of CLF into the tube model significantly modifies
its predictions. In particular,  the combination of reptation and CLF
predicts that, over a wide range of molecular weights, the
viscosity scales approximately as $N^{3.4}$ and the
diffusion coefficient as $N^{-2.4}$ with the chain length
$N$, 
in agreement with experimental findings~\cite{mcleish2002tube,
doi1983explanation}. 
The second additional relaxation mechanism is constraint release (CR),
which originates from the motion
of the surrounding chains that form the confining tube for the target
chain. It is often assumed that the relaxation of a surrounding
chain leads to a local reorganization (or a local jump) of the tube,
but the tube diameter is not perturbed.  Even though some chains
no longer contribute to the tube, others come in and impose new
constraints, such that the average structure of the melt is
conserved.
The local rearrangements of the tube due to CR events lead to a slow
Rouse-like motion of the tube, and this Rouse-like motion can
contribute to the relaxation of the chain~\cite{rubinstein2003polymer,
mcleish2002tube, rubinstein1987theory, rubinstein1988self}.  It is
worth mentioning that apart from this ``Rouse-tube'' model, the
CR mechanism has also been viewed as a process that increases the
effective tube diameter with time, \ie, leads to dynamic tube
dilation~\cite{marrucci1985relaxation,watanabe2009slow}.

Molecular dynamics simulations have proven to be efficient tools
for studying
the dynamics of entangled polymer melts and for testing
the predictions of the tube model.  For example, the tube model
predicts a series of scaling regimes for the mean-squared displacement
of monomers, $g_1(t)$, and chain center-of-mass, $g_3(t)$, with time.
In particular, it predicts an early time regime where $g_1(t)$ and
$g_3(t)$ scale as $g_1(t) \sim t^{1/4}$ and $g_3(t)\sim
t^{1/2}$, due to the restricted Rouse motion of the chain inside the tube.
These scaling regimes, or signs of the gradual development of these regimes upon increasing chain length, 
have been observed in several simulations of coarse-grained
bead-spring or chemistry-specific models~\cite{kremer1990dynamics,
putz2000entanglement, zhou2006direct, likhtman2007linear,
stephanou2010quantifying,wang2012segmental, salerno2016resolving,
hsu2016static, kempfer2019realistic, svaneborg2020characteristic,
behbahani2021dynamics, li2021dynamics}.  The chain length-dependence
of the diffusion coefficient and the zero shear viscosity
have also been investigated using simulations, and the
transition from unentangled Rouse-like behavior to entangled behavior
has been detected~\cite{kroger2000rheological,
harmandaris2003crossover, lahmar2009onset, shanbhag2020molecular,
behbahani2021dynamics, li2021dynamics}. 
Few 
simulation works have studied the single-chain dynamic structure
factor, $S(q,t)$, of entangled polymer
melts~\cite{kremer1990dynamics,putz2000entanglement, harmandaris2003crossover,
hsu2017detailed}.  This 
dynamical observable is experimentally accessible, \eg, using
the neutron spin echo technique~\cite{richter2005neutron,
gold2019direct, monkenbusch2023dynamic, kruteva2024cooperative}.  The tube model predicts a
fast partial decay of $S(q,t)$ due to local motions 
of the chain inside the tube followed by a slow long-time decay due to
escaping from the tube~\cite{de1981coherent,richter2005neutron}.
The Time-scale separation between these two processes leads to
the appearance of a plateau-like regime in the $S(q,t)$ curve, which
has been observed in molecular dynamics
simulations~\cite{hsu2017detailed}.  The viscoelastic properties of
entangled polymer melts, which have been studied in detail
experimentally~\cite{ferry1980viscoelastic, colby1987melt,
van2002evaluation, auhl2008linear} and are essential for the
applications of polymers, have also been calculated in
simulation~\cite{zhou2006direct, likhtman2007linear, hou2010stress,
tassieri2018rheo, kempfer2019realistic,
behbahani2021dynamics,li2021dynamics, liang2022multiscale,
shireen2023linear}.  In particular,  Likhtman
\etal~\cite{likhtman2007linear} calculated the shear stress relaxation
modulus, $G(t)$, of bead-spring polymer models over a rather wide
range of chain lengths, from unentangled to moderately entangled
chains.
The overall relaxation of polymer chains can be directly studied by analyzing the dynamics of the end-to-end vector, quantified through $P(t)$, the autocorrelation function of the end-to-end vector. This quantity is related to the dielectric relaxation spectra of   
 the so-called type A polymers that have dipole moments along the chain backbone~\cite{watanabe2001dielectric,abou2010rouse,glomann2011unified}.
 Despite providing direct information about the relaxation of the chains, this quantity has not been extensively studied using simulations.  
 Single-chain slip-spring models, with the capability of adjusting CR
parameters, have been used to study the effect of CR environment on
the dynamics of the end-to-end
vector~\cite{shivokhin2017understanding, read2018contour}.  More
recently, using a multiscale approach (with a multi-chain slip-spring
model at the coarsest level of description), it has been reported that
the pure reptation model can not describe the 
dynamics 
of long polyisoprene chains~\cite{li2021dynamics}.  Note that
single-chain or multi-chain slip-spring models are coarse-grained and
computationally efficient models that do not strictly prevent
chain crossings, but mimic entanglement effects by a set of springs
that restrict the lateral movements of the chain~\cite{hua1998segment,
masubuchi2001brownian, likhtman2005single, sukumaran2009modeling,
chappa2012translationally,uneyama2012multi, langeloth2013recovering,
ramirez2017multi, vogiatzis2017equation}.

The present work complements and goes beyond the previous
contributions in several aspects: (i) We provide a detailed
joint analysis of various quantities that characterize the
chain dynamics: the autocorrelation function of
the end-to-end vector, $P(t)$, the single-chain dynamic structure
factor, $S(q,t)$, the linear viscoelastic properties,
characterized by $G(t)$, and the mean-squared displacements.
This is done for a wide range of molecular weights ranging from
the unentangled to entangled regimes through long-time molecular
dynamics simulations up to the terminal time. The data of $S(q,t)$ are
also used for the calculation of non-local mobility functions, that can be used in
investigating the dynamics of structure development in heterogeneous
polymer melts.  (ii) We discuss different
relaxation processes and analyze their signatures.
To directly detect the
signatures of CLF, we mainly analyze  $P(t)$ and compare its
behavior with that of $g_1(t)$.  To detect the signatures of CR,
different dynamical observables are compared with each other.  A
central quantity in this analysis is the ``surviving tube fraction''
$\mu(t)$, a fundamental quantity in the tube model,
which denotes the fraction of the (primitive) chain that still
remains in the original tube formed at $t = 0$.  In the absence of CR
(\ie, assuming a fixed tube in space),
the tube model predicts a relation
$S(q,t) \propto G(t) \propto P(t) = \mu(t)$~\cite{de1981coherent,
watanabe2001dielectric, mcleish2002tube}  at late times $t$. It
reflects the fact that,
in the absence of CR, the behavior of these quantities should
be governed by the process of escaping from the tube, as
quantified by $\mu(t)$. We note that
the correlation between different dynamical properties has also been
investigated  in experimental works,
particularly using parallel rheology and dielectric relaxation
spectroscopy techniques~\cite{watanabe2001dielectric,
matsumiya2000comparison}.  (iii) 
We perform a detailed comparison of the
simulation results with theoretical expressions based on different
relaxation mechanisms.
The $P(t)$ and $S(q,t)$ data are
compared with the predictions of the pure reptation model
and the combination of
reptation and CLF.  The $G(t)$ data are compared with the predictions
of the Likhtman-McLeish model~\cite{likhtman2002quantitative} which
considers reptation, CLF, and CR mechanisms. The latter
discussion is similar in spirit to one in a previous study by Hou et
al.~\cite{hou2010stress}, however, 
our approach is different and the results confirm a simple relation between $G(t)$ and $P(t)$.

The manuscript is organized as follows: In the next section, we
introduce the model used for the simulations. Then, we analyze
mean-squared displacements,  dynamics of the end-to-end
vector, single-chin dynamic structure factor, and linear viscoelastic
properties, respectively. Finally, in the last section, we summarize
the results.

\section{Model}
\label{sec:model}

\begin{figure}[htb!]
    \centering
        \includegraphics[width=0.45\textwidth]{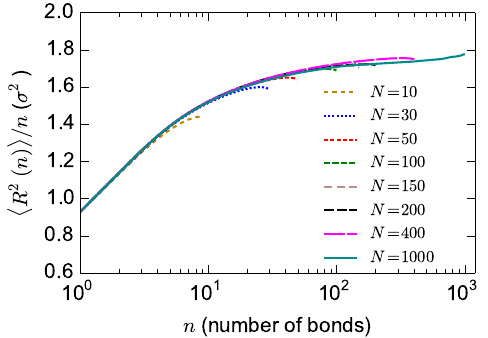}
\caption{Mean-squared internal distances of the simulated chains.
$R^2(n)$ is the mean-squared distance between two monomers
separated by $n$ bonds along a chain.}
    \label{Fig:int-dist} 
\end{figure}

Molecular dynamics simulations have been performed with the standard fully flexible
Kremer-Grest bead-spring model~\cite{kremer1990dynamics}.  In this
model, all beads have mass $m$ and purely repulsive non-bonded
interactions described by the Weeks-Chandler-Anderson
potential\cite{WCA} with characteristic length scale $\sigma$ and
prefactor $\varepsilon$. The bond-stretching interactions are
given by a FENE potential that prevents chain crossing 
and no angle-bending potential is applied.  All quantities are
expressed in units $\sigma$, $\varepsilon$, and $m$, \eg, the
basic time unit is $\tau = \sqrt{m \sigma^2/\varepsilon}$.
The bead density is equal to $\rho=0.85 \sigma^{-3}$.  Simulations
were carried out at $k_B T = 1 \varepsilon$ using a Langevin
thermostat, with bead friction $0.5 \tau^{-1}$, and a time step
$\mathrm{d}t = 0.01 \tau$ was used for integrating the equations
of motion using the LAMMPS package~\cite{thompson2022lammps}.
Simulations were performed for a wide range of chain lengths $N$,
\ie, $N =5$, $10$, $20$, $30$, $50$, $100$, $150$, $200$, $400$, and $1000$ (the results for $N =5$ and $20$ are reported in few cases).  For
these values of $N$, the simulation box contained $4800$, $2400$, $500$, $800$,
$400$, $400$, $400$, $400$, $192$ and $216$ chains, respectively.
Depending on $N$, production runs were carried out for around $10^5 \tau$
up to $10^8 \tau$ ($10^7$ to $10^{10}$ time steps).  In most cases,
the initial pre-equilibrated configurations for the simulations were
taken from previous simulations of melts of a chemistry-specific
coarse-grained model~\cite{behbahani2021dynamics}. After converting
the units and adjusting the melt density to $\rho =0.85 \sigma^{-3}$,
these systems were subjected to long-time equilibration runs.
To confirm that the resulting structures were equilibrated, we
measured 
the internal distances of the model chains, shown in
\cref{Fig:int-dist}. This figure shows $R^2(n)/n$ versus $n$, where
$R^2(n)$ is the mean-squared distance between the two monomers
separated by $n$ bonds along a chain.  For large $n$, $R^2(n)/n$ tends
to a constant value and does not have significant fluctuations, as
expected for the random-walk structure of the chains at large scales.  

Based on the size of the chains with length $N \ge 200$, we estimate
the effective segment size to be $b = 1.32 \sigma$, using
$R_\text{ee}^2 = (N-1)b^2$, where $R_\text{ee}^2$ is the mean-squared
end-to-end distance of a chain containing $N$ beads.  Also, we estimate
the Kuhn length to be $l_\text{k} = 1.81 \sigma$, using
$R_\text{ee}^2 = N_\text{k}l_\text{k}^2$ and $R_\text{max} =
N_\text{k}l_\text{k}$ where $R_\text{max} = (N-1) l_\text{b}$ is
the contour length of the chain. Here, $l_\text{b} = 0.965 \sigma$ is the average bond length between two consecutive beads along
a chain.   For  Kremer-Grest chains with length $N > 100$, the
monomeric friction coefficient has been previously calculated from the
relaxation of the large Rouse
modes,
giving $\zeta =  25 \tau^{-1}$~\cite{kremer1990dynamics,li2012nanoparticle,kalathi2014rouse,svaneborg2020characteristic}.  Also, through the primitive
path analysis~\cite{everaers2004rheology}, the entanglement length of large model chains has been
estimated to be close to $N_\text{e,ppa} =
87$\cite{moreira2015direct,hoy2009topological}. The entanglement
length is related to the  plateau modulus $G_N^0$ via $G_N^0 =
\frac{4}{5} \rho k_\text{B}T/N_\text{e}$. The value of the plateau modulus
has been estimated by Likhtman \etal~\cite{likhtman2007linear} from the
simulation of a single-chain slip-link model parameterized based on
the bead-spring model utilized in this work, giving $G_N^0 = 0.013\
\varepsilon/\sigma^3$ corresponding to $N_\text{e} = 52$. 
As we will discuss in detail in Section \cref{sec:LVE},
this value also provides a consistent description of the 
plateau modulus in the present simulations.
Given this latter estimate of $N_\text{e}$, 
the step length of the tube (or distance between entanglements along one chain),
 which can be calculated from the end-to-end distance of a subchain of length $N_\text{e}$ segments, equals $N_\text{e}b^2 = 9.52\sigma$. 
In the current work, the tube diameter, $a$, is assumed~\cite{likhtman2002quantitative,richter2005neutron} to be equal to the step length of the tube.

\section{Results and discussion}

\subsection{Mean-squared displacements}
\label{sec:msd}

\begin{figure*}[t] \centering
        \includegraphics[width=0.45\textwidth]{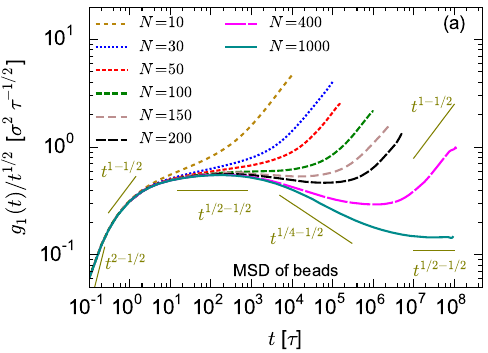}
        \includegraphics[width=0.45\textwidth]{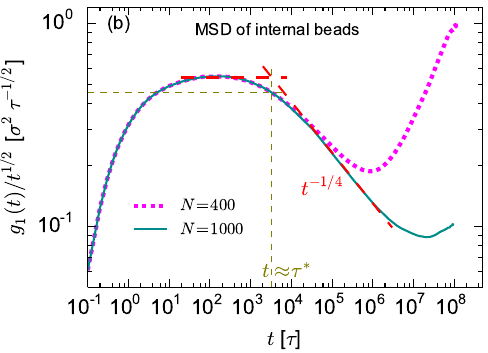}

        \includegraphics[width=0.45\textwidth]{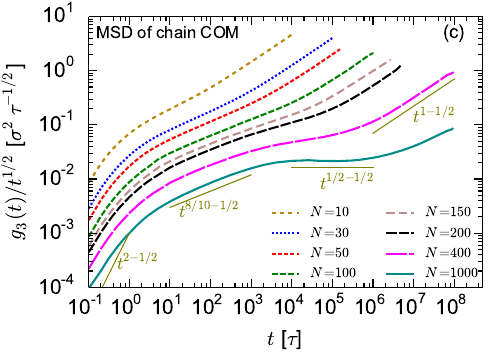}
        \includegraphics[width=0.45\textwidth]{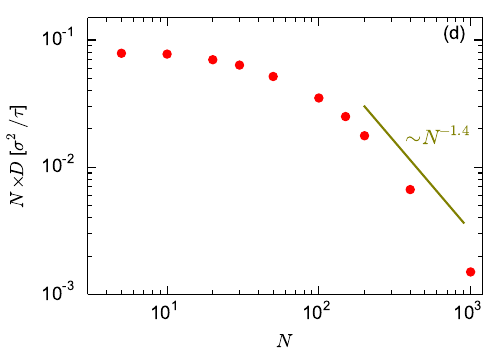}
    \caption{(a) Mean-squared displacement of all monomers normalized with $t^{1/2}$. (b) Mean-squared displacement of the internal monomers normalized with $t^{1/2}$. For $N = 400$ and $1000$, averaging has been performed over $25$ and $100$ internal monomers respectively. (c) Mean-squared displacement of the centers-of-mass of the chains divided by $t^{1/2}$. (d) $D N$ vs. $N$ where $D$ is the self-diffusion coefficient of the chains calculated from the long-time behavior of $g_3(t)$.   }
    \label{Fig:MSD} 
\end{figure*}

An initial overview of different dynamical regimes in the
system can be obtained by inspecting
the average mean-squared displacement of monomers (beads), $g_1(t)$,
and the average mean-squared displacements of the centers-of-mass of
the chains, $g_3(t)$:

\begin{align}
\begin{split}
    g_1(t) &=   \langle [{\mathbf{r}}(t) - {\mathbf{r}}(0)]^2 \rangle, \\
    g_3(t) &=   \langle [{\mathbf{r}_\text{CM}}(t) - {\mathbf{r}_\text{CM}}(0)]^2 \rangle.
\end{split}
\end{align}
Here $\langle \rangle$ denotes averaging over all target monomers/chains and 
time origins in a trajectory.

Both $g_1(t)$ and $g_3(t)$ exhibit different behaviors in unentangled
and entangled melts. For unentangled melts, the Rouse model predicts
the following scaling regimes in the limit of long
chains~\cite{doi1988theory}:
\begin{equation}
g_1(t) \sim  \left\{
\begin{array}{llr}
t^{1}\hspace*{7mm} & t<\tau_0
\hspace*{15mm} &{\text{(I$_{\text{R}}$)}}\\
t^{\frac{1}{2}}& \tau_0<t <\tau_\text{R}
&{\text{(II$_{\text{R}}$)}} \\
t^{1}& t>\tau_\text{R}
&{\text{(III$_{\text{R}}$)}}\\
\end{array} \right.,
\label{Eq:Rouse-g1}
\end{equation}     
where $\tau_0 = \zeta b^2/(3\pi^2 k_\text{B}T)$ is a monomeric time
and $\tau_\text{R} = \tau_0 N^2$ is the Rouse time of the chain.  For
entangled melts, the tube model predicts the following regimes (in
the absence of
CR)~\cite{doi1988theory,kremer1990dynamics,rubinstein2003polymer}:
\begin{equation}
\label{Eq:g1-tube}
g_1(t) \sim
\left\{
\begin{array}{llr}
t^{1} \hspace*{7mm} & t<\tau_0 \hspace*{15mm}& {\text{(I)}}\\
t^{\frac{1}{2}}& \tau_0<t<\tau_\text{e} & {\text{(II)}}\\
t^{\frac{1}{4}}& \tau_\text{e}<t<\tau_\text{R} & {\text{(III)}}\\
t^{\frac{1}{2}}& \tau_\text{R}<t<\tau_\text{d} &{\text{(IV)}}\\
t^{1}& t>\tau_\text{d} & {\text{(V)}}
\end{array} \right.,
\end{equation}     
where $\tau_\text{e} = \tau_0 N_\text{e}^2$ is the entanglement
time, $\tau_\text{R} = \tau_0 N^2$ is the Rouse time, and
$\tau_\text{d} \sim N ^{\alpha}$ is the disentanglement time of the
chain, where $\alpha \approx 3.4$ over a wide range of chain
lengths (based on the reptation and CLF
mechanisms~\cite{mcleish2002tube}).  At time scales shorter than
$\tau_\text{e}$ the chains do not feel the entanglements, and perform
unrestricted Rouse motion.  After $\tau_\text{e}$ the monomers are
constrained to move along their confining tubes.  In the range of
$\tau_\text{e} < t < \tau_\text{R}$ the chain performs restricted
incoherent Rouse motion, and in the range of $\tau_\text{R} < t <
\tau_\text{d}$, coherent diffusion inside the tube.  Because of the
random-walk structure of the tube, these exponents are half 
of the normal Rouse exponents (\cref{Eq:Rouse-g1}).  After
$\tau_\text{d}$, when the chains fully escape from their
confining tubes, $g_1(t)$ shows normal diffusive behavior.  In the
following, we will label these different time regimes as regime I-V
as indicated in \cref{Eq:g1-tube}. Different regimes in the
Rouse case are denoted I$_{\text{R}}$, II$_{\text{R}}$, and 
III$_{\text{R}}$, see \cref{Eq:Rouse-g1}.  

\cref{Fig:MSD}a shows the chain length dependence of $g_1(t)$,
calculated by averaging over all monomers, including chain ends.  To
better 
illustrate deviations from the Rouse behavior, $g_1(t)$ is divided by the Rouse slope, $t^{1/2}$ (characterizing the regimes
II$_{\text{R}}$ and II).
At very short times, 
a nearly ballistic regime ($g_1(t) \sim t^2$) is observed.
This regime reflects the inertia of monomers and is not present in
the Equations (\ref{Eq:Rouse-g1}) and (\ref{Eq:g1-tube}), which are
based on models for overdamped Brownian motion. In the Lennard-Jones
units, the transition from underdamped to overdamped motion is
expected around $t \approx 1 \tau$. At very late times, $g_1(t)$
reflects normal diffusive behavior ($g_1(t) \sim t$) with the
diffusion constant of the whole chain. At intermediate times,
different regimes are observed depending on the length of the chain.
The $g_1(t)$ of short chains ($N < 100$) always grows faster than
$t^{1/2}$ before reaching normal diffusion.  This can be
attributed to the finite length of the chains.  Indeed, the Rouse model
for short chains (finite number of Rouse modes) also predicts
exponents larger than $1/2$~\cite{behbahani2021dynamics}.  For $N \le
100$, $g_1(t)/t^{1/2}$ has positive or zero slopes before the normal diffusion regime, but for $N > 100$, an interval of negative slope (in a double logarithmic plot) gradually
emerges, which is a characteristic of entangled dynamics. With increasing chain length $N$, the negative slope becomes gradually 
more pronounced and 
approaches the prediction of the tube model, 
$g_1(t)/t^{1/2} \sim t^{-1/4}$  (regime III in
\cref{Eq:g1-tube}).

Among all monomers, the behavior of $g_1(t)$ for middle monomers
in a chain is expected to come closest to the prediction of the tube
model~\cite{kremer1990dynamics}.  Therefore, \cref{Fig:MSD}b shows separately the
function $g_1(t)$ of middle monomers for chains of length
$N = 400$ and $N = 1000$ (averaged over $25$ and $100$ middle
monomers, respectively).
As expected, the transition between scaling regimes is much
more pronounced, and
for $N = 1000$, a $t^{1/4}$ regime can clearly be seen.
Based on these data, the  early  
$t^{1/2}$ and the $t^{1/4}$ regimes intersect, approximately, at time $\tau^{\star} = 3320 \tau$ and $g_1(\tau^\star) = 26.3 \sigma^2$.
For the estimation of the intersection time, we determine the intersection 
between the two tangent lines corresponding to the $t^{1/2}$ and
the $t^{1/4}$ regimes in a double logarithmic plot as shown in
\cref{Fig:MSD}b (dashed red lines).
We note that the transition between different regimes is continuous and therefore the determination of the transition time has a degree of uncertainty. 
The transition point can be used for roughly estimating the entanglement time $\tau_\text{e}$ and the tube diameter $a$ (assuming $a^2 = N_\text{e}b^2$)~\cite{kremer1990dynamics,putz2000entanglement,likhtman2002quantitative,stephanou2010quantifying,hsu2016static,tassieri2018rheo,svaneborg2020characteristic,hou2017note}. Different expressions have been put forward in the literature.
Kremer and Grest~\cite{kremer1990dynamics} assumed $\tau_\text{e} = \tau^\star$ and $g_1(\tau^\star) = 2R_\text{g}^2(N_\text{e})= a^2/3$.
These relations give $\tau_\text{e} = 3320 \tau$ and $a = 8.9 \sigma$. 
From the values of $\tau_\text{e}$ or $a$, one can also estimate the entanglement length
$N_\text{e}$ via the relations
$a^2 = N_\text{e} b^2$ or 
$\tau_\text{e} = \tau_0 N_\text{e}^2$ 
quoted earlier. Therefore, the above values of $\tau_\text{e}$ and $a$ correspond to $N_\text{e} = 48$ and $N_\text{e} = 45$, respectively. These estimates of $N_\text{e}$ are consistent with each other and also are close to $N_\text{e} = 52$, corresponding to the plateau modulus.
Likhtman and McLeish~\cite{likhtman2002quantitative} also proposed expressions close to those mentioned above. 
More recently, Hou~\cite{hou2017note} suggested to use 
$\tau_\text{e} = 9\tau^\star/\pi$ and $g^\star = 2/(3\pi)a^2$, where $g^\star$ (which is slightly different from $g_1(\tau^\star)$ ) is taken from the intersection point of the tangent lines on the $g_1(t)$ curves (red dashed lines in \cref{Fig:MSD}b).
These relations give $\tau_\text{e} = 9516 \tau$ and $a = 12.1 \sigma$. 
These values of $\tau_\text{e}$ and $a$ correspond to $N_\text{e} = 80$ and $N_\text{e} = 84$, which are larger than the above estimates based on the suggestion of Kremer and Grest. Similar values have been reported 
previously~\cite{hou2017note}.
We summarize these rough estimates of $N_\text{e}$, calculated based on $g_1(t)$,
in Table \ref{Tab:Ne}. This table also contains the $N_\text{e}$  values calculated based on the plateau modulus and primitive path analysis, and the $N_\text{e}$  values used for the description of the simulation data based on theoretical models, as discussed in other sections.

\begin{table}[htb!]
{\footnotesize
\begin{tabular}{|ll||c||c|c|}
\hline 
\multicolumn{2}{|l||}{Quantity/method}
& $N_\text{e}$ & $a = \sqrt{N_\text{e}} b$& $\tau_\text{e} = \tau_0 N_\text{e}^2 $ \\ 
\hline
\multicolumn{2}{|l||}{Plateau modulus~\cite{likhtman2007linear}
(\cref{sec:LVE})}
&  $52$ & $9.5 \sigma$& $4.0 \cdot 10^3 \tau$ \\ \hline
\multicolumn{2}{|l||}{
Primitive path analysis~\cite{moreira2015direct,hoy2009topological}}
& $87$ &$12.3 \sigma$& $11\cdot10^3\tau$ \\
\hline
\multicolumn{5}{|l|}{$g_1(t)$ (\cref{sec:msd})}  \\ 
\cline{2-5}
 &
 $\tau_\text{e} = \tau^\star$
& 48 &  & $^*3.3 \cdot 10^3 \tau$   \\
\cline{3-5}
  \ \ \ & $a^2 =3g_1(\tau^\star)$~\cite{kremer1990dynamics} & $45$ & $^*8.9 \sigma$ & \\ 
 \cline{2-5}
 &
 $\tau_\text{e} = 9 \tau^\star/\pi$
& 80 &  & $^*9.5 \cdot 10^3 \tau$    \\
\cline{3-5}
  \ \ \ & $a^2 = 3\pi g^\star/2$~\cite{hou2017note} & 84 & $^*12.1 \sigma$ & \\ 
\hline
\multicolumn{2}{|l||}{$P(t)$ (\cref{sec:pt}), Eq.\ \ref{Eq:LM}}
&  $52$ &  &  \\ 
\hline
\multicolumn{5}{|l|}{$S(q,t)$ (\cref{sec:sqt})}  \\ 
\cline{2-5}
 &
 Pure reptation, Eq.\ (\ref{Eq:sq-tube} \& \ref{Eq:esc-rep})
& 156 & $16.5 \sigma$ &   \\
\cline{2-5}
 &
 CLF, Eq.\ (\ref{Eq:sq-tube} \& \ref{Eq:esc-clf})
& 100 & $13.3 \sigma$ &   \\
\hline
\end{tabular}
\caption{
Summary of the $N_\text{e}$ values calculated based on the plateau modulus, primitive path analysis, and monomeric mean-squared displacements ($g_1(t)$), together with the 
$N_\text{e}$ values used for the description of the measured $P(t)$ and $S(q,t)$ data based on the theoretical models. 
The corresponding values of the tube diameter, $a$, and entanglement time, $\tau_\text{e}$, are also shown in some cases.
In the case of $g_1(t)$, the stars mark the quantities estimated directly from the simulation data. 
The relations between $N_\text{e}$, $a$, and $\tau_\text{e}$ are shown in the table. We use these relations with $b=1.32 \sigma$, $\zeta = 25 \tau^{-1}$, and
$\tau_0 = \zeta b^2/(3 \pi^2k_{\text{B}} T) = 1.47 \tau$ for the
effective segment size, the monomeric friction coefficient, and the
monomeric time, respectively.}
    \label{Tab:Ne} 
}
\end{table}

Next, we consider the
chain length dependence of $g_3(t)$, shown in
\cref{Fig:MSD}c. As before, the curves are divided by $t^{1/2}$ to
better illustrate the transition between different scaling regimes. 
The tube model predicts the following regimes for the behavior of
$g_3(t)$ of entangled
chains~\cite{kremer1990dynamics,mcleish2002tube}:
\begin{equation}
g_3(t) \sim 
\begin{cases}
t^{1}\ \  \  \  \  \ t<\tau_\text{e}\\
t^{\frac{1}{2}}\ \  \  \  \  \ \tau_\text{e} < t <\tau_\text{R}\\
t^{1}\ \  \  \  \  \ t>\tau_\text{R}
\end{cases} \ \ .
\end{equation}     

The sub-diffusive $t^{1/2}$ regime (corresponding to the
$t^{1/4}$ regime of $g_1(t)$) is important because it is
not present in the $g_3(t)$ of unentangled Rouse chains.  As shown in
\cref{Fig:MSD}c, upon increasing chain length, a plateau region,
corresponding to the $t^{1/2}$ regime of $g_3(t)$,  gradually emerges
in the $g_3(t)/t^{1/2}$ curves.  For $N = 1000$, a fully
developed plateau is observed, in agreement with the theoretical
prediction.  The short-time behavior of $g_3(t)$ is also worth
attention. After the initial nearly ballistic regime, all
$g_3(t)$ curves evolve almost with $t^{0.8}$, which is different
from the theoretically expected $t^1$ scaling.  This behavior has
already been observed previously and was attributed to the
inter-molecular interactions, which are ignored in the Rouse
model~\cite{guenza2002intermolecular}.  Figure S1 in Supporting
Information (SI) shows $g_3(t)$ divided by $t^{0.8}$. In such a
presentation, the gradual development of the $t^{1/2}$ regime for the
entangled chains appears as the gradual development of a region with a
negative slope. In line with the behavior of $g_1(t)/t^{1/2}$, a
negative slope of $g_3(t)/t^{0.8}$ is seen for $N > 100$.  

\cref{Fig:MSD}d shows  $N D$ vs. $N$, where $D$ is the
self-diffusion coefficient of the chains calculated from the long-time
diffusive regime of $g_3(t)$, using $g_3(t) = 6D t$.  In the
simulation window, the behavior of $g_3(t)$ for the chain
length $N=1000$ comes close to the normal diffusion regime
($g_3(t)$ reaches values around $3 R_\text{g}^2$ during the simulation
time). For shorter chains with length $N < 1000$, normal
diffusion is clearly observed at late times. Therefore, overall,
reasonable measurements of $D$ can be performed.
Note that, besides the chain lengths given in \cref{Fig:MSD}c,  the diffusion coefficients for $N =5$ and $20$ are also included in 
\cref{Fig:MSD}d.  
The Rouse scaling, $D
\sim N^{-1}$,  is seen
in the neighborhood of the shortest chain
lengths under consideration here.
A similar behavior has been observed previously~\cite{kremer1990dynamics}. 
Note that, deviations from the Rouse behavior were
reported~\cite{behbahani2021dynamics,li2021dynamics} in simulations
of chemistry-specific unentangled melts; in those cases, the major
source of the discrepancy is the increase of the melt density with
chain lengths (in the range of small lengths).
In the present simulations, the density is independent of
$N$ and the Rouse behavior is observed for sufficiently short chains. 
For large $N$, 
the chain length dependence of $D$, 
shown in \cref{Fig:MSD}d,
is consistent with the theoretical prediction
$D \sim N^{-2.4}$, assuming a
combination of reptation and CLF.  
Pure reptation
would lead to the scaling relation $D \sim N^{-2}$.  Taking
also  CLF into account, one expects
a relation close to $D \sim N^{-2.4}$ over a wide range of
molecular weights~\cite{mcleish2002tube}. Similar scaling
relations have also been observed
experimentally\cite{lodge1999reconciliation,tao2000diffusivity}.  

\subsection{Dynamics of the end-to-end vector}
\label{sec:pt}

\begin{figure*}[!htb]
    \centering
        \includegraphics[width=0.45\textwidth]{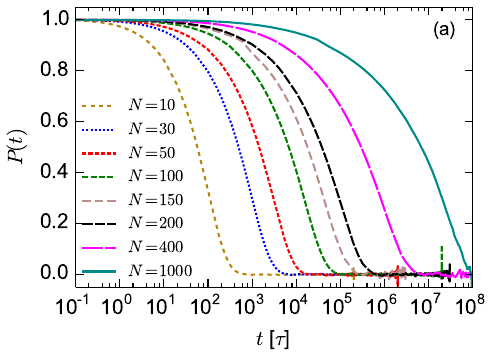}
        \includegraphics[width=0.45\textwidth]{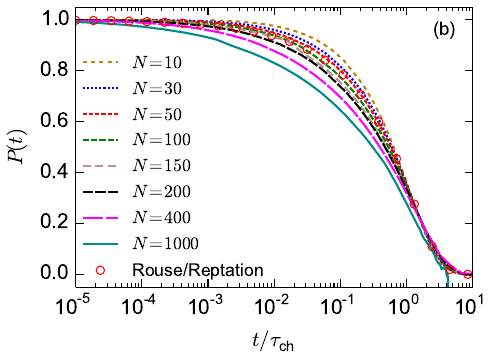}

        \includegraphics[width=0.45\textwidth]{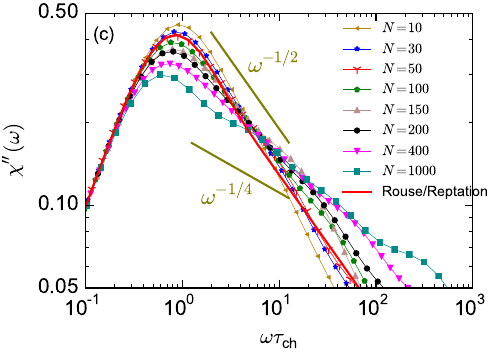}
        \includegraphics[width=0.45\textwidth]{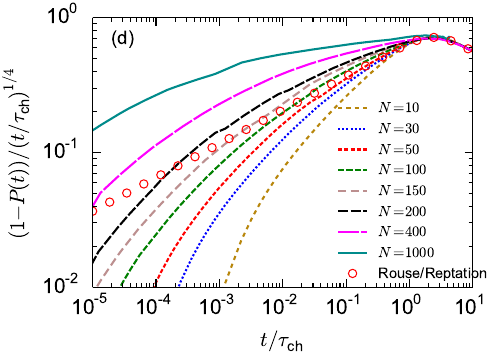}
\caption{(a) Autocorrelation function of the end-to-end
vector, $P(t)$ vs. time $t$. (b) Same data vs. 
$\tau/\tau_\text{ch}$, where $\tau_\text{ch}$ is the
mean relaxation time of $P(t)$. (c) Same data
in the frequency domain. Here the dynamic susceptibility,
$\chi^{\prime\prime}(\omega)$, is shown as a function
of $\omega \tau_\text{ch}$. (d) $(1- P(t))/(t/\tau_\text{ch})^{1/4}$
vs. $t/\tau_\text{ch}$. 
}
\label{Fig:pt} 
\end{figure*}

The overall relaxation of polymer chains can be investigated by
analyzing the dynamics of the end-to-end vector,
quantified via the autocorrelation function
\begin{equation}
    P(t) = \frac{\langle {\mathbf{R}}(t). {\mathbf{R}}(0)\rangle}{\langle {\mathbf{R}}(0). {\mathbf{R}}(0)\rangle},
\end{equation}
with ${\mathbf{R}}(t)$ being the end-to-end vector at time $t$ and $\langle \rangle$ denoting averaging over different chains and time origins in a trajectory. 
This quantity is related to dielectric relaxation 
spectroscopy measurements~\cite{mcleish2002tube,likhtman20121}.  More specifically,
the so-called dynamic susceptibility
$\chi^{\prime\prime}(\omega) = \omega \int_0^{\infty} P(t) \cos
(\omega t) \mathrm{d}t$ is expected to behave very similarly to
the dielectric loss function of the type A polymers~\cite{watanabe2001dielectric}.

The simulation results for $P(t)$ and
$\chi^{\prime\prime}(\omega)$
are shown in \cref{Fig:pt}a,b, and \cref{Fig:pt}c,
respectively. For the calculation of
$\chi^{\prime\prime}(\omega)$, we fitted $P(t)$ with a series of
exponential functions (Debye modes), $P(t) = \sum_i a_i
\exp(t/\tau_i)$, giving $\chi^{\prime\prime}(\omega) = \sum_i a_i
\omega \tau_i/(1 + \omega^2 \tau_i^2)$. In \cref{Fig:pt}b,c, we
have rescaled $t$ and $\omega$ with the mean relaxation
time, defined as $\tau_\text{ch} = \int_0^\infty P(t) \ud t$, to make
the curves comparable. 

The Rouse model and the pure reptation model both predict a
similar relation for $P(t)$\cite{doi1988theory}: 
\begin{equation}
  P_\text{\tiny Rouse/rep}(t) = \frac{8}{\pi^2}
    \sum_{p,\text{odd}}^{\infty} \frac{1}{p^2} \exp(-\frac{p^2 t}{\tau_1}),
\label{Eq:pt-rept} 
\end{equation} 
where $\tau_1$ is the longest relaxation time of the chain, \ie, $\tau_1 =
\tau_\text{R}$ in the Rouse model and $\tau_1 = \tau_\text{d}$  in the
reptation model. According to this equation, the mean
relaxation time should be given by
$\tau_\text{ch}^\text{\tiny Rouse/rep} = (\pi^2/12) \tau_1$, and
all curves for $P(t)$ should collapse onto a single master curve 
if plotted against $t/\tau_\text{ch}$. Furthermore, the quantity
$(1-P(t))$ should scale as 
\begin{equation}
\label{Eq:pt-rept2}
\begin{aligned}
  1-P(t) &= 
     \frac{8}{\pi^2}\sum_{p, \text{odd}}^{\infty}
     \frac{1}{p^2}(1-\ue^{-p^2t/\tau_1})
   \\ &\approx 
      \frac{4}{\pi^2}\int_0^{\infty} \!\!\!\!\! \ud p \:  
         \frac{1}{p^2} (1-\ue^{-p^2 t/\tau_1})
       = \frac{4}{\pi^{3/2}} \: (\frac{t}{\tau_1})^{1/2}
 \end{aligned}
\end{equation}
for $t \ll \tau_\text{1}$, which corresponds to a scaling behavior
$\chi^{\prime\prime}(\omega) \sim \omega^{-1/2}$ of the dynamic
susceptibility at $\omega \gg \tau_{\text{1}}^{-1}$.

In order to test these predictions, we first compare the 
curves for $P(t)$ vs.\ $t/\tau_{\text{ch}}$ in \cref{Fig:pt}b
with the prediction of \cref{Eq:pt-rept} (symbols). The curves
clearly do not collapse. The simulation data for
$P(t)$ at chain lengths $N = 50$ and $N = 100$ are close  to the prediction of \cref{Eq:pt-rept}, however, 
for shorter chains the $P(t)$ curves are less stretched than this relation and for longer chains
deviations from this relation become gradually more pronounced and the curves for $P(t)$ become more and more stretched. This behavior is consistent with the gradual stretching of the dielectric relaxation curves with increasing chain length, as detected experimentally~\cite{boese1990molecular}.  
Furthermore, $P(t)$ can be written as the sum of the autocorrelations of the odd Rouse modes of the chain~\cite{doi1988theory,likhtman20121} and the gradual stretching of $P(t)$ is consistent with the gradual stretching of the autocorrelation of the Rouse modes with increasing chain length, as observed previously~\cite{kalathi2014rouse,hsu2017detailed}.

\cref{Fig:pt}c shows the corresponding
$\chi^{\prime\prime}(\omega)$ curves alongside with the prediction
of \cref{Eq:pt-rept} which predicts a peak at $\omega
\tau_{\text{ch}}\approx 1$ followed by the power law decay mentioned
above, $\chi''(\omega) \sim \omega^{-1/2}$ on the high frequency side.
Consistent with the trend observed in \cref{Fig:pt}b, for short chains,
the curves seem close to the theoretical prediction; 
for large chain lengths $N$, however, the peak shifts to smaller 
$\omega \tau_\text{ch}$-values and broadens, and the
slope of $\log(\chi^{\prime \prime}(\omega))$  vs.\ $\log \omega$
at high frequencies gradually deviates from $-1/2$ and approaches
$-1/4$. 
These observations are consistent with the results of
dielectric relaxation spectroscopy experiments on entangled
melts~\cite{watanabe2001dielectric,abou2010rouse,glomann2011unified}. Although, we should note that 
experiments on highly entangled melts show the appearance of two scaling regimes on the high-frequency side of the relaxation peak: $\omega^{-1/4}$ scaling at high frequencies (far from the peak frequency) and $\omega^{-1/2}$ scaling regime at lower frequencies, close to the peak frequency~\cite{watanabe2001dielectric,abou2010rouse,glomann2011unified}. In the range of chain length studied here, 
we do not still see the $\omega^{-1/2}$ scaling regime for entangled chains. 
A qualitatively similar chain-length-dependent shape of the dielectric relaxation curve has also recently been observed in multiscale
simulations of polyisoprene melts of different
lengths~\cite{li2021dynamics}.

Finally, \cref{Fig:pt}d tests the scaling behavior of
$(1-P(t))$ vs. $t/\tau_\text{ch}$ in a double logarithmic plot, where $(1-P(t)$ is scaled with $(t/\tau_\text{ch})^{1/4}$, again showing
the prediction of \cref{Eq:pt-rept} for comparison.  Whereas the
theory curve (symbols) features the expected $t^{1/2}$ scaling for all
times below $t < \tau_{\text{ch}}$, none of the simulation curves
follows this behavior strictly. 
$(1-P(t))$ has a rich structure and its slope is both time and $N$ dependent. 
Particularly, upon increasing chain length, a region corresponding to $(1 - P(t)) \sim t^{1/4}$ gradually emerges. 
Such a scaling regime has been previously observed in the slip-spring simulations as a plateau region in the plot of $-t^{3/4}\partial P/\partial t$ vs. $t$~\cite{read2018contour}.

\begin{figure}[!htb]
    \centering
        \includegraphics[width=0.45\textwidth]{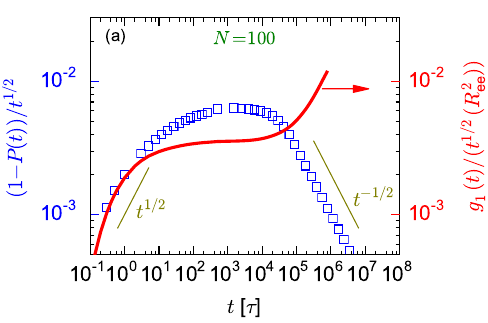}
        
        \includegraphics[width=0.45\textwidth]{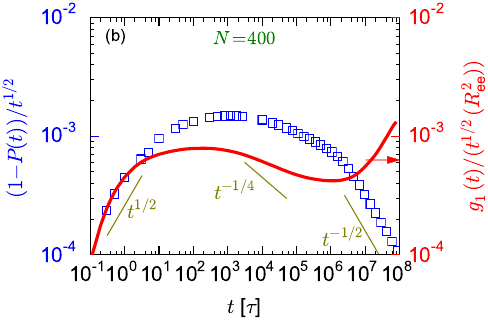}
        
        \includegraphics[width=0.45\textwidth]{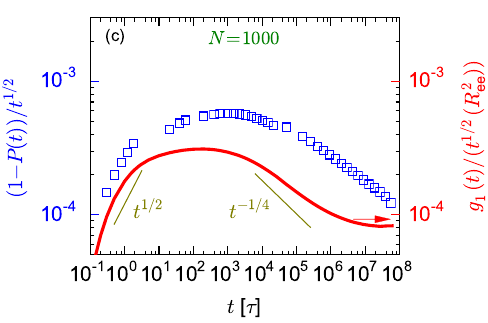}
    \caption{ $(1- P(t))/t^{1/2}$ together with rescaled $g_1(t)/t^{1/2}$ 
      (averaged over all monomers) vs.\ time $t$. 
    Panels (a), (b), and (c) show the curves for 
    $N = 100$, $N = 400$, and $N = 1000$, respectively.    
    }
    \label{Fig:pt-g1t} 
\end{figure}

As will be discussed below, the $\omega^{-1/4}$ and $t^{1/4}$ behaviors of
$\chi^{\prime\prime}(\omega)$ and $(1-P(t))$, respectively, can be explained by CLF. This behavior originates from the restricted Rouse motion of the chain inside the tube in the interval of $\tau_\text{e}<t<\tau_\text{R}$, which also leads to 
the subdiffusive $t^{1/4}$ scaling of $g_1(t)$ (regime III of \cref{Eq:g1-tube})~\cite{rubinstein2003polymer}.
Before discussing this regime in more detail, we analyze the structure of $P(t)$ by plotting 
$(1-P(t))$ together with  $g_1(t)$,  for $N = 100$,  $400$, and $1000$, in \cref{Fig:pt-g1t}.
 In this figure, both functions are normalized by $t^{1/2}$.
Furthermore, to facilitate the comparison, $g_1(t)/t^{1/2}$ is divided by $R^2_\text{ee}$, such that both functions have similar values at short times.
$(1-P(t))$ approximately follows the time-dependence of $g_1(t)$. At short times, before feeling entanglements ($t < \tau_\text{e}$), $(1-P(t))$ first scales approximately with $t$ and then with $t^{1/2}$;
this latter scaling regime is the one expected for a free Rouse chain (as \cref{Eq:pt-rept2} shows).  
After these short-time regimes, the $(1-P(t))/t^{1/2}$ of sufficiently long chains (similar to their $g_1(t)/t^{1/2}$) exhibits a negative slope which tends to $-1/4$ upon increasing $N$ (\ie, $(1- P(t))\sim t^{1/4}$ corresponding to regime III in \cref{Eq:g1-tube}).
For $N = 100$ (and shorter chains), 
the negative slope of $(1-P(t))/t^{1/2}$ before the terminal time is negligible.
At late times ($t>\tau_\text{d}$, when $g_1(t)$ has normal diffusion) $P(t)$ tends to zero and $(1-P(t))/t^{1/2} \sim 1/t^{1/2}$.
In the range of chain length studied here, corresponding to $g_1(t) \sim t^{1/2}$ in the interval of $\tau_\text{R} < t < \tau_\text{d}$,  a clearly separated regime of $(1-P(t))\sim t^{1/2}$ (in that time range) is not observed.
For longer chains, where the decay of $P(t)$ until $\tau_\text{R}$ becomes small and the separation between $\tau_\text{R}$ and $\tau_\text{d}$ becomes large, it is expected that a regime of  $(1-P(t))\sim t^{1/2}$ in the range of 
$\tau_\text{R} < t < \tau_\text{d}$ is observed.  
Such a regime corresponds to
scaling regime $\chi^{\prime\prime}(\omega) \sim \omega^{-1/2}$ of
the dynamic susceptibility, which has indeed been observed
experimentally~\cite{watanabe2001dielectric,
abou2010rouse,glomann2011unified} in the range $\tau_\text{d}^{-1} <
\omega < \tau_\text{R}^{-1}$ as mentioned earlier.

We continue with a discussion
of the $(1-P(t))\sim t^{1/4}$ regime in terms of  CLF.
For this, we make use of a 
central quantity of the tube model, the   
``surviving tube fraction'', $\mu(t)$,  which is the fraction of the (primitive) chain that is still trapped in the original tube formed at $t = 0$.
When some parts of the chain escape from the original tube (formed at $t = 0$), they become orientationally decorrelated from the orientation of the chain at $t = 0$, and only the orientations of the segments that are still trapped inside the tube remain correlated with that at $t = 0$. Therefore, in the absence of CR (because the argument assumes a fixed tube in space), $P(t)$ is equal to $\mu(t)$~\cite{doi1988theory,mcleish2002tube,watanabe2001dielectric}.
Note that this argument ignores the decay of $P(t)$ due to the very short time Rouse motion of the chain inside the tube, before feeling entanglements ($t < \tau_\text{e}$); for long chains, this motion leads to a very small decay of $P(t)$.
For pure reptation $\mu(t)$ is expressed by \cref{Eq:pt-rept} (\ie, it equals $P_\text{rept}(t)$). However, pure reptation does not take into account the restricted Rouse motion of the chain in the tube in the range of $\tau_\text{e} < t < \tau_\text{R}$.  
The effect of the restricted Rouse motion on $\mu(t)$ is discussed in terms of CLF in the literature~\cite{doi1988theory,rubinstein2003polymer,likhtman2002quantitative}.
CLF leads to the escape of a portion of the segments from the tube, which has been quantified via the following simple functional form~\cite{likhtman2002quantitative}:
\begin{equation}
    \mu(t) = 1 - \frac{C}{Z}(\frac{t}{\tau_\text{e}})^{1/4},\ \ \  t<\tau_\text{R},
\label{Eq:clf}
\end{equation}
where $Z = N/N_\text{e}$ and $C = 1.5$ according to Likhtman and McLeish~\cite{likhtman2002quantitative}. 
The above $t^{1/4}$ scaling of the escaped segments (\ie, $1 - \mu(t)$) originates from the restricted Rouse motion of the chain~\cite{rubinstein2003polymer}, which also leads to 
the $t^{1/4}$ scaling of $g_1(t)$.
Considering the equality of $\mu(t)$ and $P(t)$ in the absence of CR (for $t > \tau_\text{e}$), the  emergence of  $(1 - P(t)) \sim t^{1/4}$ for long chains, as suggested by \cref{Fig:pt}d, is consistent with the above relation. 
It is also worth mentioning a comment 
about the effect of CR on $P(t)$. The end-to-end correlation of chains in monodisperse linear
polymer melts is dominated by reptation and CLF, and the effect
of CR decreases with increasing chain length.  According to
the ``Rouse-tube'' model of CR, the tube behaves as a Rouse chain of
$Z = N/N_\text{e}$ segments. The average local relaxation time of each segment
is the disentanglement time of a surrounding chain, which scales with
$Z^{3.4}$, and the overall relaxation time of the tube due to CR
events scales with~\cite{mcleish2002tube} $Z^2 \cdot Z^{3.4} =
Z^{5.4}$.  For large values of $Z$, the CR relaxation time of the tube
is much longer than the disentanglement time and hence $P(t)$ of long chains is not
expected to be largely affected by the CR process.  The effect
of CR on the $P(t)$ of monodisperse linear polymers has also been
studied experimentally for $Z \approx 4$ up to $Z \approx 35$ by
comparing the relaxation of chains in quasi monodisperse melts
with that in blends containing
very long chains~\cite{matsumiya2013dielectric}.  It has been reported
that the CR mechanism accelerates the dielectric relaxation
(corresponding to relaxation of the end-to-end vector),
however, the effect decreases with increasing $N$.  Suppression of CR,
via blending with very long chains, has been reported to increase the
dielectric relaxation time around $35\%$ for $Z \approx 20$ and around
$22\%$ for $Z \approx 35$~\cite{matsumiya2013dielectric}; these values correspond to the relaxation times
of around $9\%$ and $6\%$ longer
chains.  
Based on these considerations, we do not consider CR and compare the measured $P(t)$ curves with the following analytical relation, proposed by Likhtman and McLeish~\cite{likhtman2002quantitative}, for the 
evolution of $\mu(t)$ due to CLF at short times and  reptation at later times:
\begin{equation}
\begin{aligned}
        \mu(t) &= \int_{\varepsilon^*(Z,\tau_\text{e})}^{\infty} \frac{0.306}{Z\tau_\text{e}^{1/4}\varepsilon^{5/4}}\exp(\varepsilon t)\mathrm{d}\varepsilon \\ 
        &+ \frac{8\Tilde{G}_\text{f}(Z)}{\pi^2} \sum_{p, \text{odd}}^{p^*(Z)} \frac{1}{p^2}\exp(-\frac{p^2 t}{\tau_\text{df}(Z,\tau_\text{e})}). 
        \label{Eq:LM}
\end{aligned}
\end{equation}
Among others, this equation
has been used for describing the experimental dielectric relaxation
curves 
of entangled polymer
melts~\cite{glomann2011unified,pilyugina2012dielectric}. 
The first
term of \cref{Eq:LM} is the contribution of CLF and dominates up
to times of the order of $\tau_\text{R}$. This term is
responsible for the scaling $(1 - \mu(t)) \sim t^{1/4}$ (see \cref{Eq:clf}). The second term is the contribution
of reptation which dominates
at times longer than $\tau_\text{R}$. For large values of $Z$, this
term leads to a scaling $(1-\mu(t))\sim t^{1/2}$ up to the terminal time (see also \cref{Eq:pt-rept2}). 
The time $\tau_\text{df}(Z)$ is the disentanglement time in the
presence of CLF, and $\Tilde{G}_\text{f}(Z)$ is a dimensionless
plateau modulus.  Explicit expressions for the relations
between these quantities and $\varepsilon^*$ and $p^*$ can be
found in ~\textcite{likhtman2002quantitative}. The
quantities $Z = N/N_\text{e}$ and $\tau_\text{e}$ are the only input
variables in  \cref{Eq:LM}. In our analysis, we set $N_\text{e} = 52$, which is the value extracted from the plateau modulus (see \cref{sec:model} and \cref{Tab:Ne}) and calculate
$\tau_\text{e}$ from
$\tau_\text{e} = \tau_0 N_\text{e}^2 = 1.47 N_\text{e}^2$ (see
\cref{sec:msd}). 

\begin{figure}[!htb]
    \centering
        \includegraphics[width=0.45\textwidth]{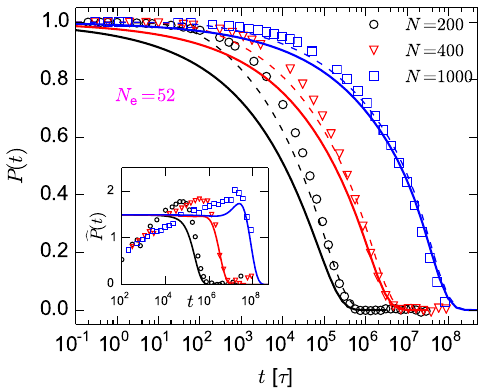}
  
\caption{(a) The $P(t)$ data for
$N = 200$, $400$, and $1000$ (symbols) compared with the prediction for $\mu(t)$ of the Likhtman-McLeish model,
\cref{Eq:LM}. The solid lines are the results of the original model
(\cref{Eq:LM}) using $N_\text{e} = 52$ (estimated from the plateau modulus). The dashed lines show the
results of the modification proposed by Hou et
al.~\cite{hou2010stress} where the upper bound of the integration
in \cref{Eq:LM} is replaced by $(\alpha^4 \tau_\text{e})^{-1}$.
$\alpha = 0.35$ is used here. Inset:  same data in the presentation adopted by Likhtman and McLeish~\cite{likhtman2002quantitative}; here 
$\widehat{P}(t) = -4Z\tau_\text{e}^{1/4} t^{3/4} \partial P/\partial t$.
}
    \label{Fig:LM} 
\end{figure}

\cref{Fig:LM} compares the simulation data (symbols) for $P(t)$ 
with the results of \cref{Eq:LM} (solid lines) for
chain lengths $N=200, N=400,$ and $N=1000$ in two different
presentations.
For $N = 200$, \cref{Eq:LM} does not satisfactorily describe $P(t)$.
This is to be expected, since \cref{Eq:LM} predicts an
initial scaling regime $(1-P(t))\sim t^{1/4}$, which is not yet
developed for such mildly entangled chains. 
However, the quality of fitting increases with increasing $N$, and for
$N = 1000$, a fair agreement between the measured $P(t)$ and
\cref{Eq:LM} is obtained, if one excludes the 
early regimes I and II of unrestricted Rouse motion, which do not  contribute
significantly to the decay of $P(t)$. 
To show the effect of $N_\text{e}$ on the prediction of the model (\cref{Eq:LM}), in
SI, Figure S2, the results of using 
$N_\text{e,ppa}=87$ obtained from the primitive path analysis, are shown.
With $N_\text{e,ppa}=87$, the deviation between the model and the simulation data is large, as compared to the results of using $N_\text{e}=52$.

In their study of the viscoelastic properties of polymer melts,  
Hou
\etal~\cite{hou2010stress} suggested replacing the upper bound of
integration in the first term of \cref{Eq:LM} by $(\alpha^4
\tau_\text{e})^{-1}$, where $\alpha$ is a fitting parameter, and
rescaling $\mu(t)$ such that it starts
from $1$ at $t = 0$.   
This eliminates the main contribution to
the decay of $\mu(t)$ at times shorter than around
$(\alpha^4 \tau_\text{e})$.
Here, we also show the results of this modified equation in \cref{Fig:LM} (dashed lines; $\alpha = 0.35$).
The behavior
for short chains improves albeit at the expense of introducing
a fitting parameter.  
The output is sensitive to the value of $\alpha$ as the bound of integration changes with $\alpha^4$.

\begin{figure}[!htb]
    \centering
        \includegraphics[width=0.45\textwidth]{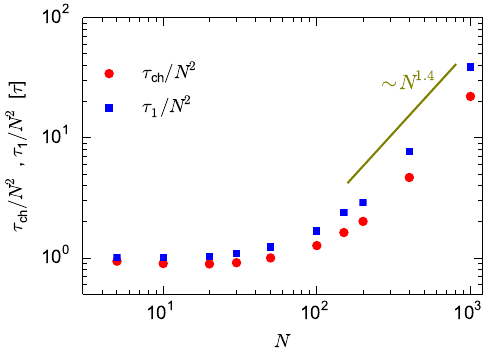}
\caption{Average relaxation time $\tau_\text{ch} = \int_0^\infty
P(t) \ud t$ and maximum relaxation time $\tau_1$ of $P(t)$ vs. chain
length $N$.}
    \label{Fig:pt-tau} 
\end{figure}

We close this section by discussing the average and the longest
relaxation times of $P(t)$, $\tau_\text{ch}$ and $\tau_1$,
respectively. They are shown in \cref{Fig:pt-tau}. The
relaxation times are divided by $N^2$. The time $\tau_1$ has been
estimated from the relaxation time of the slowest exponential mode
used for the fitting of $P(t)$.  In the range of chain lengths
$N$ studied here, $\tau_\text{d}/\tau_\text{ch}$ increases with
increasing $N$, 
due to the fact that the $P(t)$ curves are gradually more
stretched.
The $N$ dependence of of $\tau_1$ shows a gradual change from the
Rouse behavior ($\tau_1 \sim N^2$) to entangled behavior. As
mentioned above,  based on the reptation and CLF mechanisms, we
expect $\tau_\text{1} \sim N^{\alpha}$ with $\alpha \approx 3.4$
over a wide range of molecular
weights~\cite{doi1983explanation,mcleish2002tube}. This is consistent
with the behavior observed here.
Similar scaling behavior was previously observed for polyisoprene melts via multiscale simulations~\cite{li2021dynamics} and also dielectric relaxation spectroscopy experiments~\cite{adachi1985dielectric,matsumiya2013dielectric}.

\subsection{Single-chain dynamic structure factor}
\label{sec:sqt}

\begin{figure}[!htb]
    \centering
        \includegraphics[width=0.45\textwidth]{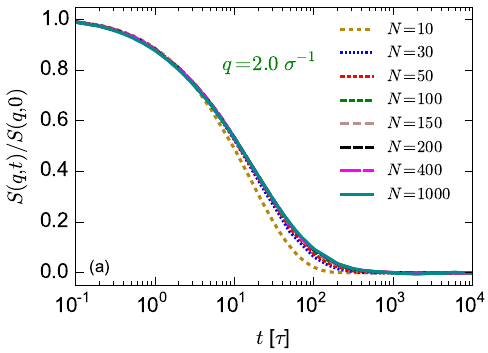}

        \includegraphics[width=0.45\textwidth]{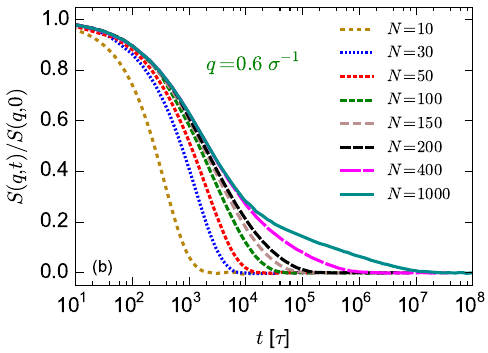}

\caption{(a) Normalized dynamic structure factor $S(q,t)/S(q,0)$
vs.  time $t$ (a) at $q = 2 \sigma^{-1}$, corresponding to a
length scale much smaller than the tube diameter, and (b) $q =
0.6 \sigma^{-1}$ corresponding to a length scale comparable to the tube
diameter ($2\pi/a = 0.66\ \sigma^{-1}$).
}
    \label{Fig:sq1} 
\end{figure}

In this section, we discuss the dynamic structure factor of single chains,
which characterizes the dynamics of the chain
at different length scales and time scales.  It is defined as 
\begin{equation} 
S(q,t) =  \frac{1}{N} \sum_{m,n}^{N} 
  \langle \exp\{i\mathbf{q} \cdot [\mathbf{r}_m(t) - \mathbf{r}_n(0)]\}\rangle,
\end{equation} 
where the sum is taken over the monomers of a chain. 
\cref{Fig:sq1}a shows the chain length dependence of $S(q,t)/S(q,0)$ at $q = 2.0 \sigma^{-1}$.
This $q$ value corresponds to the length scales much smaller than the tube diameter ($2\pi/a = 0.66\ \sigma^{-1}$) and also smaller than, although comparable to, the size of the smallest studied chain ($2\pi/R_\text{ee}^{N=10} \approx 1.75\ \sigma^{-1}$).
Based on the Rouse model, $S(q,t)/S(q,0)$ is independent of the chain length on length scales smaller than the chain size, $q \gg
2\pi/R_\text{ee}$\cite{doi1988theory}. Also, entangled chains are expected to behave similarly at length scales much smaller than tube diameter ($q \gg 2 \pi/a$) where the chains do not yet feel the entanglement constraints.
The results shown in \cref{Fig:sq1}a are consistent with this theoretical expectation and at $q = 2.0 \sigma^{-1}$, $S(q,t)/S(q,0)$ for all studied chains are similar. 

\cref{Fig:sq1}b shows the $S(q,t)/S(q,0)$  at $q = 0.6 \sigma^{-1}$.
This $q$ value corresponds to a length scale comparable to the tube diameter, however larger than the end-to-end distances of the chains with length $N \le 50$.
At length scales larger than the chain size ($q \ll 2\pi/R_\text{ee}$), $S(q,t)$ is controlled by the self-diffusion coefficient of the chains, $S(q,t)/S(q,0) = \exp(-Dq^2 t)$; this explains the chain-length dependence of $S(q,t)/S(q,0)$ at $q = 0.6 \sigma^{-1}$ for $N \le 50$.  
For $N \ge 100$, the short-time behavior of $S(q,t)$ is almost chain length independent, but, at long times $S(q,t)$ develops a pronounced $N$-dependent shoulder.
Previously, Hsu and Kremer~\cite{hsu2017detailed} studied the chain length dependence of $S(q,t)/S(q,0)$ for long entangled chains and they also observed similar behavior. 
Qualitatively similar behavior has also been observed in neutron spin echo experiments~\cite{kruteva2024cooperative}.
According to the tube model, $S(q, t)$ of entangled polymers, in length scales larger than the tube diameter and smaller than the chain size, has a two-step decay: the first step originates from the local motion of the chain inside the tube, and the second step comes from the escape of the chain from the tube (creep)~\cite{de1981coherent}.
The characteristic time of the local process is chain length independent whereas that of the creep process 
is disentanglement time, $\tau_\text{d}$, which strongly sales with $N$ ($\sim N^{3.4}$)~\cite{de1981coherent}.
The trend observed in \cref{Fig:sq1}b is in qualitative agreement with the expected behavior of $S(q, t)$ based on the tube model.

\begin{figure}[!htb]
    \centering
        \includegraphics[width=0.45\textwidth]{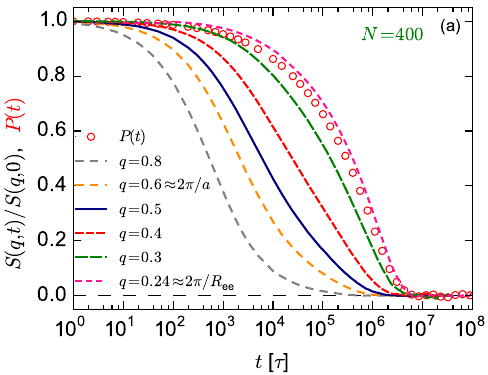}

        \includegraphics[width=0.45\textwidth]{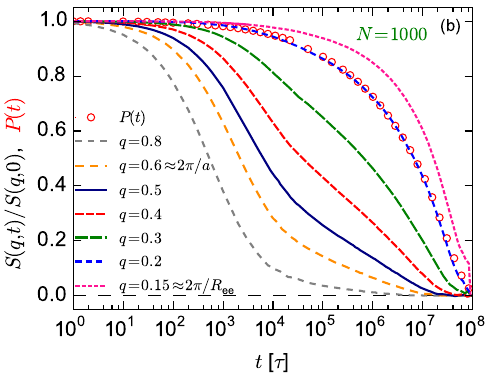}
    
        \includegraphics[width=0.45\textwidth]{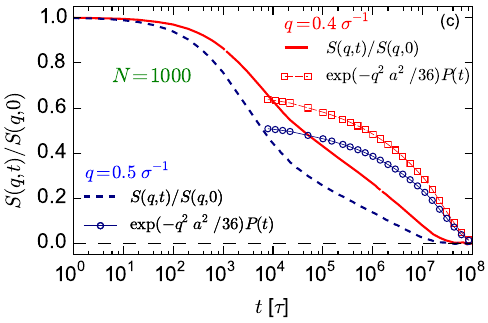}

\caption{(a,b) Normalized dynamic structure factor $S(q,t)/S(q,0)$ vs. time
$t$ for different $q$-vectors at chain length $N=400$ (a) and $N=1000$
(b). Also shown for comparison is $P(t)$, the autocorrelation 
function of the end-to-end vector.
(c) $S(q,t)/S(q,0)$ of $N = 1000$ at $q = 0.4 \sigma^{-1}$ and $q = 0.5 \sigma^{-1}$  together with the corresponding $\exp(-q^2a^2/36)P(t)$ functions.  }

    \label{Fig:sq-qs} 
\end{figure}

\cref{Fig:sq-qs}a-b shows the $q$-dependence of $S(q,t)/S(q,0)$ for $N = 400$ and $N = 1000$. 
This figure covers the range from $q = 2\pi/R_\text{ee}$ to $q = 0.8 \sigma^{-1}$. The latter $q$ value corresponds to a length scale rather smaller than the tube diameter ($2\pi/a = 0.66 \sigma^{-1}$), and even at this length scale, $S(q,t)$ exhibits a shoulder. 
With decreasing $q$, the height of the shoulder increases.
As mentioned above, based on the tube model, in the range of
$2\pi/R_\text{ee} \ll q \ll 2\pi/a $ the long-time behavior (\ie, the shoulder) of
$S(q,t)$ is governed by the creep process which has a $q$-independent
characteristic time of
$\tau_\text{d}$ (in the absence of CR)~\cite{doi1988theory,de1981coherent}.
This expectation is based on the following argument~\cite{de1981coherent,doi1988theory}: 
$S(q,t)$ is dominated by the pair of monomers for which
$|\mathbf{r}_m(t) - \mathbf{r}_n(0)| < 2\pi q^{-1}$. At time $t$, a
monomer that has escaped from the tube is separated from monomers
inside the original tube (formed at time $0$)  by distances comparable to the chain size. Thus, for $q$ values corresponding to length
scales much smaller than the chain size, pairs containing an escaped
particle do not contribute to $S(q,t)$, and  the
long-time behavior of $S(q,t)$ should be controlled by 
(and be proportional to)
 the fraction of monomers which are still confined in
the original tube, \ie, the surviving tube fraction, $\mu(t)$.  This argument assumes that the tube is fixed in space and therefore
neglects CR.
In the absence of CR, we have $\mu(t) = P(t)$ and the
argument thus predicts $S(q,t)/S(q,0) \propto P(t)$ at late times. 
The proportionality constant, which is the amplitude of the creep process is predicted to be
$\exp(-q^2a^2/36)$~\cite{de1981coherent,mcleish2002tube,richter2005neutron}. Therefore, the tube model, in the absence of CR, predicts that in the range of  
$2\pi/R_\text{ee} \ll q \ll 2\pi/a $ the long-time behavior of $S(q,t)$ can  be described by  $S(q,t)/S(q,0) = \exp(-q^2a^2/36) \: P(t)$.
Therefore, it is also expected that in that $q$ range, the shoulder of $S(q,t)/S(q,0)$ has a $q$-independent characteristic (decay) time.
We do not observe this for the chain lengths studied here, and the times where $S(q,t)/S(q,0)$ vanishes in \cref{Fig:sq-qs} do not coincide over a $q$ range.
More explicitly, the proportionality of $S(q,t)$ and $P(t)$ is not observed. 
 \cref{Fig:sq-qs}c
shows $S(q,t)$ at
$q=0.4\sigma^{-1}$ and 
$q=0.5\sigma^{-1}$  together with the corresponding $\exp(-q^2a^2/36) \: P(t)$ curves for $N = 1000$.
One can see that 
the long-time decay of $S(q,t)/S(q,0)$ is
significantly faster than that of $P(t)$. 
This signals the
importance of CR in the relaxation of $S(q,t)$, at least for the chain lengths studied here.
We note
that, to our knowledge, the long-time regimes of $S(q,t)$ and
$P(t)$ have not been compared experimentally, most probably because of
the limited time window of neutron spin echo experiments.

\begin{figure}[b]
    \centering
        \includegraphics[width=0.45\textwidth]{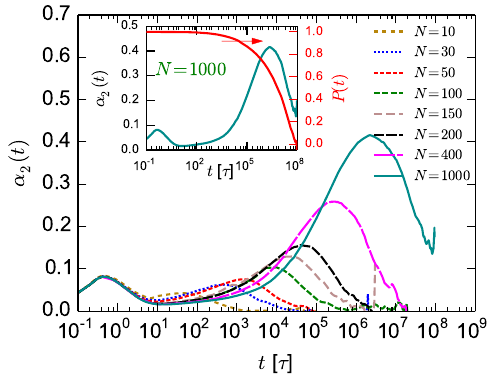}
\caption{Non-Gaussian parameter for the displacement of
the beads, $\alpha_2(t)$, versus time for different chain
lengths $N$ as indicated.  Inset:
$\alpha_2(t)$ of $N = 1000$ together with its $P(t)$, the
autocorrelation function of the end-to-end vector.}
    \label{Fig:alpha} 
\end{figure}

The
argument sketched above for discussing the long-time behavior of $S(q,t)$, relies on the assumption that there are two
classes of monomers: Monomers that are still trapped inside the tube
after a time $t$ whose displacements are restricted, and monomers that
have escaped from the tube and have larger displacements. This
heterogeneity can be visualized by calculating
the second-order non-Gaussian parameter $\alpha_2(t) = 3\langle \Delta
r ^4 \rangle/5\langle \Delta r ^2 \rangle^2 -1$, where, $\langle
\Delta r ^2 \rangle =g_1(t)$ is the mean-squared displacement of
the monomers.
\cref{Fig:alpha} shows the $N$ dependence of $\alpha_2(t)$. At
short times, $\alpha_2(t)$ exhibits a very small $N$-independent peak,
originating from the interparticle caging effect. At later times,
$\alpha_2(t)$ features a second peak which becomes higher with
increasing $N$.  This behavior has been observed previously for
different model chains~\cite{guenza2014localization,goto2021effects}. 
For comparison, we have also calculated $\alpha_2(t)$ for the ideal Rouse chains
with lengths $N = 100$ and $N = 1000$, through 
Brownian dynamics simulations of  ensembles of
non-interacting Gaussian chains (data not shown).  The resulting curves are almost
chain-length independent when plotted against $t/\tau_R$; they also
feature a peak, which can be attributed to the motion of the end monomers, but the maximum is around $0.05$ (almost
equal to the maximum of $\alpha_2(t)$ for $N = 30$ in
\cref{Fig:alpha}). In contrast, the peak values of
$\alpha_2(t)$ in the model bead-spring chains are much larger and
increase with increasing chain length. 
This supports the picture
discussed above regarding the existence of two groups of monomers with
small and large displacements for long chains. 
However, for the studied chain lengths, the heterogeneity of displacements is not still large enough to see the proportionality of $S(q,t)/S(q,0)$ and $P(t)$.
Note that, similar to the Rouse chains, the monomers with large displacements are the monomers near the chain ends because the end monomers escape the tube before the central monomers. However, as discussed above, the intra-chain dynamical heterogeneity of entangled chains is qualitatively different from the corresponding property of the Rouse chains.
The inset of \cref{Fig:alpha} shows $\alpha_2(t)$ together with
$P(t)$ for $N = 1000$.  
The development of the peak of $\alpha_2(t)$
coincides
with the decay of $P(t)$ and $\alpha_2(t)$ reaches its
maximum before the final decay of $P(t)$.

\begin{figure*}[!htb]
    \centering
        \includegraphics[width=0.45\textwidth]{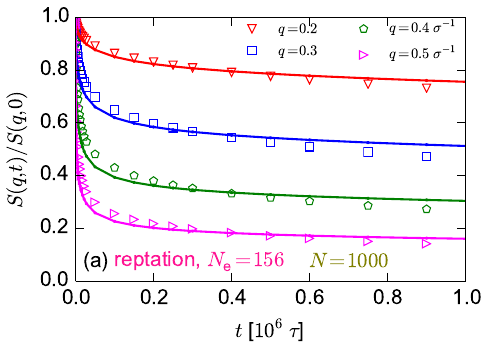}
        \includegraphics[width=0.45\textwidth]{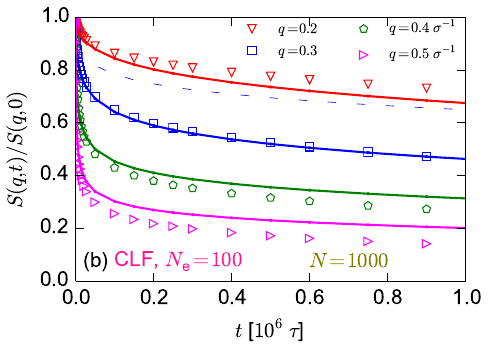}
        \includegraphics[width=0.45\textwidth]{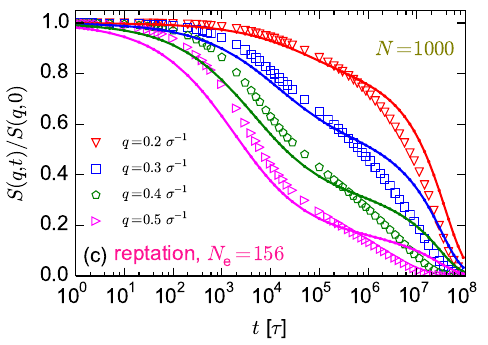}     
        \includegraphics[width=0.45\textwidth]{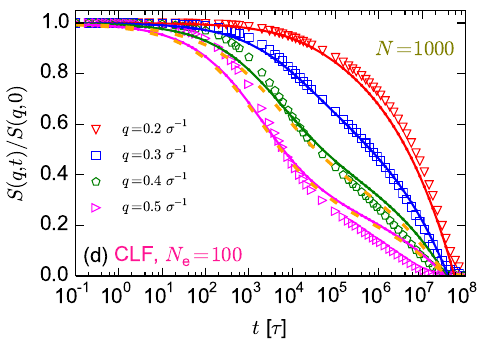}     

\caption{Simulation results for  $S(q,t)/S(q,0)$
of $N = 1000$ for different values of $q$ (symbols) together with
theoretical predictions (solid lines):
panels (a)
and (c) show, in linear and semi-logarithmic scales, the predictions for $S(q,t)/S(q,0)$ based on
the pure reputation model (\cref{Eq:sq-tube} and \cref{Eq:esc-rep}) 
with $N_\text{e} = 156$ corresponding to $a = 16.5 \sigma$.
Panels (b) and (d) show the predictions based on the CLF model (\cref{Eq:sq-tube} and \cref{Eq:esc-clf}) with $N_\text{e} = 100$ corresponding to $a= 13.3 \sigma$.
The thin dashed line in (b) shows the result obtained with $N_\text{e}=52$
(see \cref{Tab:Ne}) at $q=0.3\ \sigma^{-1}$ for comparison. 
Yellow dashed lines in (d) show the results obtained with
\protect\cref{Eq:esc-muLM}. 
}
    \label{Fig:sq-models} 
\end{figure*}

We proceed with a more quantitative comparison of the simulation results with
predictions of the tube model.
Accounting for the processes of the local motion of the chain
inside the tube (local reptation) and the escape of the chain from the tube, 
the following expression has been proposed for
$S(q,t)$~\cite{de1981coherent,mcleish2002tube,richter2005neutron}:
\begin{equation}
\begin{aligned}
\frac{S(q,t)}{S(q,0)} &= [1 - \exp(-\frac{q^2 a^2}{36})]S^\text{loc}(q,t)\\
&+ \exp(-\frac{q^2 a^2}{36})S^\text{esc}
(q,t).
\label{Eq:sq-tube}
\end{aligned}
\end{equation}
The first term describes local motion and
is calculated using
\begin{equation}
S^\text{loc}(q,t) = 
\exp(\frac{t}{t_\text{loc}})
  \text{erfc}(\sqrt{\frac{t}{t_\text{loc}}}),
\end{equation}
where $t_\text{loc} = 12\zeta/(k_\text{B} T b^2 q^4)$ is the
characteristic time of the local process, which is  chain
length independent and scales with $q^{-4}$ (it has Rouse behavior).
The second term, $S^\text{esc}(q,t)$, accounts for the escape of
the chain from the tube.
The following expression has been calculated
by assuming the pure reptation process as the only escape mechanism of the chain~\cite{doi1988theory}:
\begin{equation}
\begin{aligned}
        S^\text{esc}(q,t) &= \frac{1}{S(q,0)} \sum_{p=1}^{\infty}[\frac{2\nu N}{\alpha_p ^2(\nu^2 + \alpha_p^2 + \nu)}\\
        &\times\sin^2{\alpha_p} \exp(-\frac{4 t \alpha_p^2}{\pi^2
        \tau_{d}})],
    \label{Eq:esc-rep}
\end{aligned}
\end{equation}
where $\nu = q^2 N b^2/12$  and $\alpha_p$ are the solutions of
$\alpha_p \tan \alpha_p = \nu$, and $\tau_\text{d} = \zeta N^3
b^2/(k_\text{b}T \pi^2 N_\text{e})$ is the disentanglement time.  
Since pure reptation does not provide a proper description of the
escaping process (as discussed in section \cref{sec:pt}), the
following relation has been calculated for $S^\text{esc}(q,t)$ at $t <
\tau_\text{R}$ due to CLF (assuming that CLF dominates escaping
process at this time range)~\cite{wischnewski2002molecular}:
\begin{equation}
\begin{aligned}
     S^\text{esc}(q,t) &= \frac{1}{S(q,0)}\frac{N}{2 \nu^2}[2
    \nu + \mathrm{e}^{-2\nu} + 2\\
    &- 4\nu s(t) - 4\mathrm{e}^{-2\nu s(t)} + \mathrm{e}^{-4\nu
    s(t)}].
    \label{Eq:esc-clf}
\end{aligned}
\end{equation}
Here $s(t) = \psi(t)/2$ and $\psi(t) = 1-\mu(t) =
1.5(t/\tau_\text{e})^{1/4}/Z$ is the fraction of segments that has
escaped from the tube due to CLF. 
At $t<\tau_\text{R}$, the surviving tube fraction $\mu(t) = 1- \psi(t)$ which is identical to \cref{Eq:clf}.

For comparing the above analytical expressions with the simulation results, we treat $N_\text{e}$ as the only fitting parameter. 
\cref{Eq:sq-tube} has $a$ as an explicit input; 
for a given $N_\text{e}$, we calculate $a$ using $a^2 = N_\text{e}b^2$.   
The expression for $S^\text{esc}(q,t)$ due to CLF (\cref{Eq:esc-clf}) has $N_\text{e}$ and $\tau_\text{e}$ as explicit inputs. We calculate $\tau_\text{e}$ for a given $N_\text{e}$ using $\tau_\text{e}= \tau_0 N_\text{e}^2$.
\cref{Fig:sq-models} shows the measured $S(q,t)$ for $N = 1000$, together with the theoretical fits in linear and logarithmic presentations.
On the linear scale, they are shown up to times of the order of (but smaller than) the Rouse time of chains. 
Panels (a) and (c) show the results of the pure reptation model.
The fit gives the value $N_\text{e}=156$ for
the entanglement length (corresponding to $a = 16.5\ \sigma$), which is much larger than 
$N_\text{e}=52$ calculated from the plateau modulus. 
The origin of this discrepancy is the shortcoming of the pure reptation process for the description of the creep process. Here, the ignorance of additional relaxation mechanisms is reflected in much larger apparent tube diameters than expected.
Such behavior has also been experimentally observed~\cite{wischnewski2002molecular}. 

Panels (b) and (d) of \cref{Fig:sq-models} present the results of the CLF model, fitted on the simulation data with $N_\text{e} = 100$ (corresponding to $a = 13.3\ \sigma$).
\cref{Eq:esc-clf} is valid for $t<\tau_\text{R}$, however, in panel (d) it is plotted up to the terminal time, although for fitting only the interval $t<\tau_\text{R}$  is considered. 
The linear presentation, where the times on the order of $10^6 \tau$ can be easily seen, might give the impression that the pure reptation model better fits the simulation results than the CLF model; however, the logarithmic presentation reveals that this is not the case.
The CLF model fits the $S(q,t)$ of $N = 1000$ better than the reptation model, even after $\tau_\text{R}$; however, the applicability of \cref{Eq:esc-clf} for $t>\tau_\text{R}$ should not be generalized. As illustrated in \cref{Fig:pt}d, in the case of  $1000$-bead chains, the slope of $(1-P(t))$ vs. $t$ stays close $1/4$ (\ie, the prediction of  CLF) up to the terminal time (this would not be the case for longer chains at $t>\tau_\text{R}$).
The value of $N_\text{e} = 100$ used for fitting the CLF model on the simulation results is significantly smaller than the value calculated from the pure reptation model. This shows the importance of CLF in the process of escaping from the tube (creep process).
However, $N_\text{e} = 100$ is still considerably larger than 
$N_\text{e} = 52$ of the plateau modulus.
This is consistent with \cref{Fig:sq-qs}c which shows a discrepancy between the long-time behaviors of $P(t)$ and $S(q,t)$. 
As shown in \cref{Fig:LM}, the $P(t)$ of chain with length $N = 1000$ can be fairly described by assuming $N_\text{e} = 52$; since $S(q,t)$ decays significantly faster than $P(t)$, 
a larger apparent entanglement length is needed 
for the description of $S(q,t)$ based on the CLF model.
This signals the       
importance of the CR mechanism, which is ignored in the CLF model, for the relaxation of the $S(q,t)$ of $N = 1000$.
To assess the sensitivity of the analytical relation to $N_\text{e}$, \cref{Fig:sq-models}b also shows the result for $S(q,t)$ obtained with $N_\text{e} = 52$ at $q = 0.3\ \sigma^{-1}$ (thin dashed line); it significantly deviates from the simulation data.
The $S(q,t)$ spectra of $N = 400$  are also compared with the theoretical models in Figure S3.
Like the results of $N = 1000$, the CLF model fits the
simulation data better than the reptation model.
The values of $N_\text{e}$ used for fitting the pure reptation and CLF models to $S(q,t)$ for $N = 400$ are close (have around $10\%$ difference) to the ones used in the case of $N = 1000$.

Generally, for $q \gg 2\pi/R_\text{ee}$, $S^\text{esc}(q,t)$ can be approximated by the surviving tube fraction
$\mu(t)$~\cite{de1981coherent,doi1988theory}, as explained earlier using a simple argument in the context of \cref{Fig:sq-qs}. 
Consistent with this expectation, 
in the range of $q \gg 2\pi/R_\text{ee}$, $S^\text{esc}(q,t)$ of the reptation model (\cref{Eq:esc-rep})
can be approximated with $S^\text{esc}(q,t) = \mu_\text{rep}(t) = P_\text{rept}(t)$ which is calculated using \cref{Eq:pt-rept}~\cite{doi1988theory,richter2005neutron}.
Also, in this $q$ range, $S^\text{esc}(q,t)$ of the CLF model (\cref{Eq:esc-clf}) reduces to $S^\text{esc}(q,t) \approx \mu(t) = 1- \psi(t)$ for $t < \tau_\text{R}$ (to see this relation, consider that in $q \gg 2\pi/R_\text{ee}$ range, $\nu$ is large and $S(q,0)\approx N/\nu$). 
The above relation can be generalized to times larger than $\tau_\text{R}$, up to the terminal time, using
\begin{equation} 
S^\text{esc}(q,t) = \mu^\text{LM}(t), \ \ \ q \gg 2\pi/R_\text{ee}, 
\label{Eq:esc-muLM} 
\end{equation}
where $\mu^\text{LM}(t)$ is taken from the
Likthman-McLeish model (\cref{Eq:LM}) which considers CLF (same as $\mu(t) = 1- \psi(t)$) and also the long-time reptation process.
The dashed yellow lines in \cref{Fig:sq-models}d 
show the results of \cref{Eq:esc-muLM} for 
sufficiently large 
$q > 2 \pi/R_\text{ee}$; for $N = 1000$, they are almost 
identical to the results of the \cref{Eq:esc-clf}.  
However, we again note that in general, \cref{Eq:esc-clf} is applicable up to $t < \tau_\text{R}$, while \cref{Eq:esc-muLM} is valid up to the terminal time.

\begin{figure}[tb]
    \centering
        \includegraphics[width=0.45\textwidth]{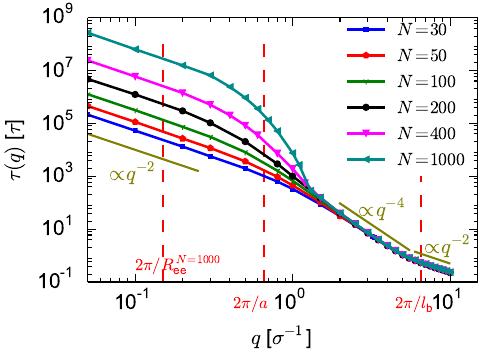}

\caption{
Effective relaxation time of $S(q,t)/S(q,0)$ as calculated from its integral
vs. wave vector $q$.}
    \label{Fig:tau_q} 
\end{figure}

After the comparison of $S(q,t)$ with theoretical predictions of
the tube model, we will now consider two quantities of special interest
that can be derived from $S(q,t)$: The spectrum of effective relaxation
times $\tau(q)$ and the closely related wave-vector dependent mobility 
function $\Lambda(q)$.

We define the effective relaxation time $\tau(q)$ in analogy to
$\tau_\text{ch}$ (see \cref{Fig:pt-tau}), via the
integral\cite{mantha2020bottom} $\tau(q)=\int_0^\infty S(q,t)/S(q,0)
\: \ud t$. 
\cref{Fig:tau_q} shows $\tau(q)$ vs.  $q$ for different chain
lengths. 
On the smallest scale, where length scales are smaller than the bond length ($q > 2\pi/l_\text{b}$), the beads
do not feel the connectivity to other beads; in this
regime, $\tau(q) \sim q^{-2}$ corresponding to the scaling behavior of normal diffusion. This length scale regime corresponds to the time scale regime I of \cref{Eq:g1-tube}.  
In the range $2\pi/a \ll q \ll 2\pi/l_\text{b}$, the beads feel the
connectivity along the chain, however, they still do not
experience entanglement constraints (corresponding to regime II of \cref{Eq:g1-tube}). In this $q$
range, $\tau(q) \sim q^{-4}$ which is the scaling behavior of the Rouse
chains on scales smaller than the chain size\cite{doi1988theory}.
On length scales that are much smaller
than the tube diameter ($q \gg 2\pi/a$), $\tau(q)$ is almost chain
length independent. 
Upon decreasing $q$ and approaching length scales comparable to $a$
(but still smaller than $a$), the values of $\tau(q)$
for entangled chains increase rapidly and become chain length
dependent.  As shown in \cref{Fig:sq1}b, 
this behavior originates from
the gradual development of chain-length-dependent shoulders in
$S(q,t)$ of entangled chains.  After the steep increase of $\tau(q)$
around $a$, the slope of $\log(\tau(q))$ vs.  $\log(q)$ decreases
significantly in the range $2\pi/R_\text{ee} < q < 2\pi/a$.  
On these length scales, the decay of $S(q,t)$ results from two relaxation processes
with two relaxation times as discussed above, and $\tau(q)$
results from the contributions of both, but
it is dominated by the slow 
creep process.  Finally, on scales $q <
2\pi/R_\text{ee}$, the chains show diffusive behavior with
$\tau(q) \sim q^{-2}$. 

The non-local 
mobility function $\Lambda(q)$ is used in dynamical density
functional theory calculations of the kinetics of the concentration
fluctuations in heterogeneous polymer melts. In fact, 
the first attempts 
to analyze $S(q,t)$ theoretically for entangled polymer melts
were motivated by the wish to calculate such
non-local mobility functions~\cite{de1980dynamics,pincus1981dynamics}.
The dynamics of concentration fluctuations in a heterogeneous melt, \eg,
a blend, is usually described using the following generalized diffusion
equation~\cite{de1980dynamics,muller2005incorporating}:  
\begin{align}
\begin{split}
\frac{\partial \phi(\mathbf{r})}{\partial t}  &= -\text{div} \mathbf{J}\\ 
\mathbf{J}(\mathbf{r}) &= - \int \mathrm{d}\mathbf{r'} \Lambda(\mathbf{r},\mathbf{r'}) \nabla \mu(\mathbf{r'})\ ,
\end{split}	
\end{align}
where $\Lambda(\mathbf{r},\mathbf{r'})$ is a non-local mobility
coefficient that connects flux at point $\mathbf{r}$ to the driving
force at point $\mathbf{r'}$.  
If the
system is translationally invariant with
$\Lambda(\mathbf{r},\mathbf{r'}) = \Lambda(\mathbf{r}-\mathbf{r'})$,
this equation translates to $J(q) = -\Lambda(q) \mathrm{i} q \: \mu(q)$ in Fourier representation.
We have recently proposed to calculate $\Lambda(q)$ via
$\Lambda(q) = {S(q,0)}/(k_\text{B}Tq^2N\tau(q))$,  where 
$S(q,0)$ is the single chain form factor and $\tau(q)$ is the
relaxation time of $S(q,t)/S(q,0)$ defined
above~\cite{mantha2020bottom}. 

\begin{figure}[!htb]
    \centering
        \includegraphics[width=0.45\textwidth]{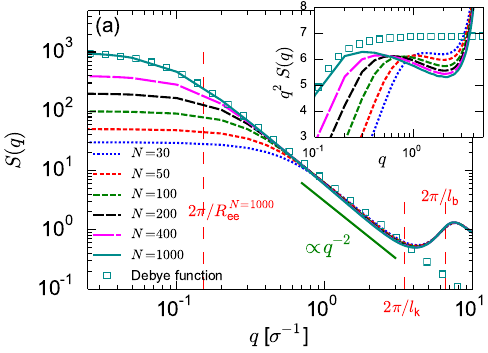}
    
        \includegraphics[width=0.45\textwidth]{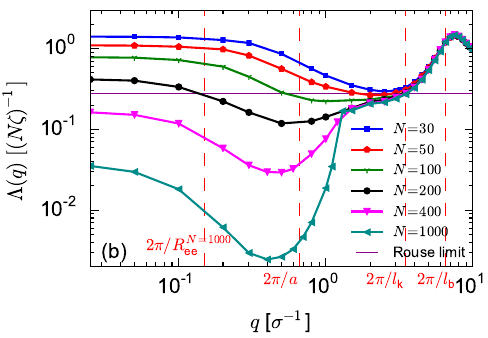}

        \includegraphics[width=0.45\textwidth]{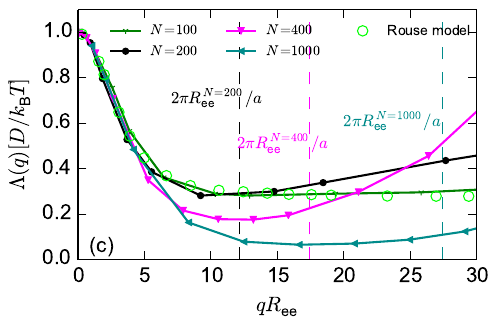}
\caption{(a) Single-chain form factor, $S(q)=S(q,0)$, for chains of
different lengths. The Debye function,  corresponding to the $S(q)$ of a
Gaussian chain, is also shown for $N = 1000$.  (b) Non-local mobility
function, $\Lambda(q)$, in units of $[(N\zeta)^{-1}]$ for chains of
different lengths.  (c) Non-local mobility function, $\Lambda(q)$, in
units of $[D/k_\text{B}T]$ vs. $qR_\text{ee}$. The symbols show
$\Lambda(q)$ as calculated from the Rouse model~\cite{schmid2020dynamic} for comparison.
     \label{Fig:mobili}} 
\end{figure}

\cref{Fig:mobili}a shows $S(q)$ ($=S(q,0)$) , the single-chain form factor (or
single-chain structure factor), for chains of different 
lengths. Also shown for comparison is the form factor of a Gaussian
chain of length $N=1000$ (symbols), which is given by the Debye
function~\cite{doi1988theory}, $S(q)/N = 2/x^2 (\exp(-x) + x - 1)$ with
$x = (qR_\text{g})^2$.
The Debye function cannot describe the behavior of the measured
$S(q)$ on small length scales, where the local chain structure is
important, but it approximately describes its behavior
at length scales larger than the Kuhn length, $l_\text{k} \approx 1.81\
\sigma$.  However, a closer look reveals detectable deviations
between  $S(q)$ and the Debye function even at length scales larger than
$l_\text{k}$, 
for all $N$.  They
can be seen in the inset of \cref{Fig:mobili}a which shows $q^2S(q)$
vs. $q$ (Kratky plot); in this presentation, the Debye function
exhibits
a plateau, which is not seen in the simulation data.  Such
deviations have been previously observed for different polymer
models~\cite{wittmer2007polymer,hsu2016static,li2021dynamics} and have
been attributed to 
the interplay of the chain connectivity and incompressibility of the
melt~\cite{wittmer2007polymer}.

\cref{Fig:mobili}b shows $\Lambda(q)$ in units of
$[(N\zeta)^{-1}]$, where $\zeta = 25 \tau^{-1}$ is the monomeric
friction coefficient as determined for long chains, see
\cref{sec:model}.  In these units, the curves for $\Lambda(q)$
coincide for all chain lengths at large wavevectors $q$.
The behavior of $\Lambda(q)$ reflects the trends observed in
\cref{Fig:tau_q} for $\tau(q)$.  For Rouse chains, $\tau(q) \sim
q^{-4}$ and $S(q)/N \sim q^{-2}$ in the range of $q\gg
2\pi/R_\text{ee}$. Accordingly, 
$\Lambda(q) = {S(q)}/(k_\text{B}Tq^2N\tau(q))$ is constant in this
$q$ range. Based on the Rouse model, one expects this constant to be
given by~\cite{schmid2020dynamic} 0.279, and this is indeed seen in
the data in \cref{Fig:mobili}b.
Entangled chains ($N > 100$) exhibit Rouse behavior in
the range of $2\pi/a \ll q < 2\pi/l_\text{K}$.  Upon decreasing $q$
and approaching length scales comparable to the tube diameter,
entanglement effects lead to a rise of $\tau(q)$ (see
\cref{Fig:tau_q}) 
and, consequently, a decrease of $\Lambda(q)$ for $N > 100$.  

The large-wavelength behavior of $\Lambda(q)$ is best evaluated in
a different presentation, which uses chain dimensions and chain
diffusion coefficient as natural units.  \cref{Fig:mobili}c shows the mobility
functions in units of $[D/(k_\text{B}T)]$, where $D$ is the diffusion 
constant of the chain, as a function of $qR_\text{ee}$.  The plots are
shown in the range $0 < qR_\text{ee}< 30$, which is usually relevant for
calculations of concentration fluctuations.  For comparison,
the prediction of $\Lambda(q)$ as calculated from the Rouse
model~\cite{schmid2020dynamic} is also shown (symbols). The simulation data for $\Lambda(q)$ at chain length $N = 100$
are very close to the result for the Rouse model.  Ideal Rouse chains
have a single length scale ($R_\text{ee}$) and their $\Lambda(q)$ is
chain length independent when presented as a function of
$qR_\text{ee}$. This is not the case for entangled melts in which the
tube diameter also plays a role; this difference in behavior
becomes apparent for $q R_\text{ee} > 2 \pi$. Here, $\Lambda(q)$ drops to
lower values and then rises again. However, in the range of $q <
2\pi/R_\text{ee} $, the chains show normal diffusive behavior and
$S(q,t)/S(q,0) \approx \exp(-D q^2 t)$. In this range the expression
for $\Lambda(q)$ simplifies to $\Lambda(q) [D/(k_\text{B}T)] \approx
S(q)/N$ which can be approximated with the chain length independent Debye function (shown in
\cref{Fig:mobili}a).  

\subsection{Linear viscoelastic properties}
\label{sec:LVE}

Finally, we discuss the shear stress relaxation modulus, which is
the basic linear viscoelastic property of a polymer melt. It can
be calculated from the autocorrelation function of shear stresses at
equilibrium:
\begin{equation}
    G(t) = \frac{V}{k_\text{B}T}
    \langle \sigma_{\alpha \beta}(t) 
    \sigma_{\alpha \beta}(0)\rangle,\ \ \  \alpha \neq \beta,
    \label{Eq:Gt}
\end{equation}
where $V$ is volume and $\sigma_{\alpha \beta}(t)$ ($\alpha, \beta \in
\{x, y, z \}$) is a shear component of the instantaneous stress
tensor.  We have used the multiple-tau
correlator algorithm~\cite{ramirez2010efficient} 
to calculate this correlation function, and to improve the
statistics, we have averaged
over different orientations of the coordinate
system~\cite{ramirez2010efficient}. 

\begin{figure}[t]
    \centering
    
        \includegraphics[width=0.45\textwidth]{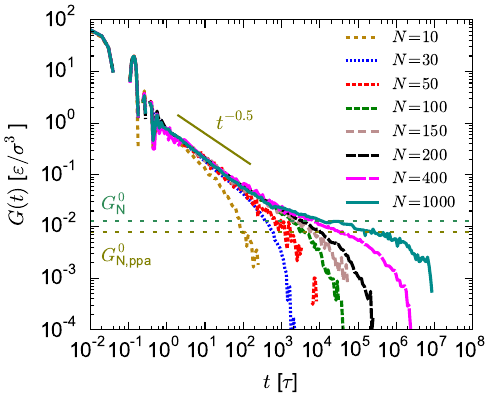}
    
    \caption{Shear stress relaxation modulus, $G(t)$, vs. time $t$
    for different chain lengths as indicated.
       The dashed lines mark two values of the plateau modulus, 
    $G^0_\text{N}$: $G^0_\text{N} = 0.013 \varepsilon/\sigma^3$ 
    corresponding to $N_\text{e} = 52$,  which appears to be a reasonable 
    estimate of the plateau modulus, and 
    $G^0_\text{N,ppa} = 0.0078 \varepsilon/\sigma^3$ corresponding to 
    $N_\text{e,ppa} = 87$ (calculated from primitive path analysis) 
    which underestimates the plateau modulus (see text).}
    \label{Fig:Gt} 
\end{figure}

The results 
are shown in \cref{Fig:Gt}.  
Generally, before the chain relaxation processes, the $G(t)$ of a polymer melt exhibits the signatures of bond vibrations (appearing as short-time oscillations) and $\alpha$-relaxation (appearing as a drop in $G(t)$ after a glassy plateau at sufficiently low temperatures)~\cite{behbahani2024local}. For the current model, $\alpha$-relaxation is very fast (it takes place around the short-time peak of
the non-Gaussian parameter $\alpha_2(t)$ in \cref{Fig:alpha}) and it is merged with the bond-vibration process. 
After these short-time processes, chain relaxation processes take
place.  The Rouse model predicts $G(t) \sim t^{-1/2}$ for $t <
\tau_\text{R}$.  Entangled chains also exhibit this Rouse scaling
before feeling entanglement constraints ($t <
\tau_\text{e}$).  After the $N$-independent entanglement time
$\tau_\text{e}$, entangled chains are constrained in their confining
tubes and for sufficiently long chains, $G(t)$ flattens and
develops a plateau. Finally, at the
disentanglement time
$\tau_\text{d}$, 
the
chains escape from the tubes which leads to the final sharp decay of
$G(t)$. 
In the range of chain lengths studied here, the plateau regime is not yet fully developed, but 
entanglement effect manifests in the emerging
$N$-dependent shoulders in the $G(t)$ curves.

The plateau modulus is related to the
entanglement length via
$G^0_\text{N} = (4/5) G(\tau_\text{e}) = (4/5)\  \rho k_\text{B}
T/N_\text{e}$, where $\rho$ is the number density of monomers. The
horizontal dashed lines in \cref{Fig:Gt} mark two estimates
of $G^0_\text{N}$ (see also \cref{sec:model} and \cref{Tab:Ne}):
(i) $G^0_{\text{N}} = 0.013 \varepsilon/\sigma^3$ corresponding to
$N_\text{e} = 52$, which
was calculated by Likhtman \etal\ for a slip-link model parameterized based on the Kremer-Grest chains
~\cite{likhtman2007linear}, and 
(ii) $G^0_{\text{N,ppa}} = 0.0078 \varepsilon/\sigma^3$ calculated
from $N_\text{e,ppa} = 87$ measured through primitive path
analysis~\cite{hoy2009topological,moreira2015direct}.
Unfortunately, a direct visual evaluation of the above estimates
for $G^0_\text{N}$ based on the simulation data is not possible, since the
curves for $G(t)$ do not exhibit a well-defined plateau with slope
zero. However, we can 
evaluate
the self-consistency of the different estimates
by
using the relation $G^0_\text{N}=(4/5) G(\tau_\text{e})$ 
to extract $\tau_\text{e}$ from the $G(t)$ curves (for a given $N_\text{e}$) in \cref{Fig:Gt}, 
and comparing this value with that obtained via
the relation $\tau_\text{e} = \tau_0 N_\text{e}^2$.
The estimate (i), $G^0_\text{N}=0.013
\varepsilon/\sigma^3$, yields
$\tau_\text{e} \approx 2.9 \cdot 10^3 \tau$
for chain length $N = 400$ and 
$\tau_\text{e} \approx 4 \cdot 10^3 \tau$ 
for $N=1000$; these values are very close to the Rouse time $N_\text{e} = 52$ monomers,  $\tau_0 N_\text{e}^2 \approx 4\cdot 10^3 \tau$.
However, the estimate (ii), $G^0_\text{N,ppa}=0.0078 \varepsilon/\sigma^3$
gives $\tau_\text{e} \approx 83 \cdot 10^3 \tau$ for $N=1000$
and $\tau_\text{e} \approx 15 \cdot 10^3 \tau$ for $N=400$; these values of $\tau_\text{e}$ are strongly $N$-dependent and also both are 
larger than $\tau_0 N_\text{e,ppa}^2 = 11\cdot 10^3 \tau$.
Thus, the estimate $N_\text{e}=52$ leads to a more consistent
overall picture, and $G^0_\text{N}=0.013 \varepsilon/\sigma^3$ appears 
to be a reasonable estimate of the plateau modulus. 
Further discussion of the value of the plateau modulus is provided below.
We note that we can also invert this argument and use the self-consistency condition to estimate $N_\text{e}$ and $G^0_\text{N}$ from the simulation data for $G(t)$, which would also give $N_\text{e} \approx 52$.
Previously Hsu and Kremer~\cite{hsu2016static} measured the $G(t)$ of the melts of long semiflexible bead-spring chains; they reported that the entanglement length determined from the primitive path analysis is consistent with the plateau modulus of the measured $G(t)$ curves. Therefore, $N_\text{e,ppa}$
is not generally inconsistent with the plateau modulus.

\begin{figure}[!htb]
    \centering
    
        \includegraphics[width=0.45\textwidth]{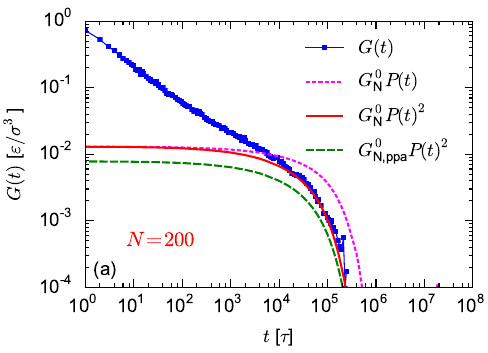}
     
        \includegraphics[width=0.45\textwidth]{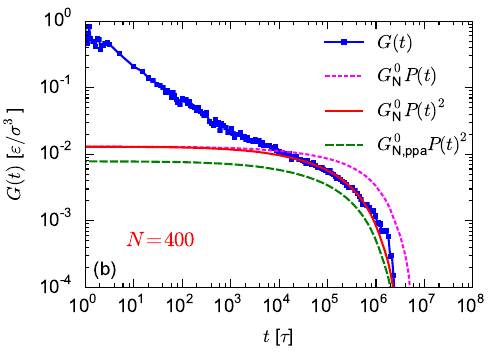}

        \includegraphics[width=0.45\textwidth]{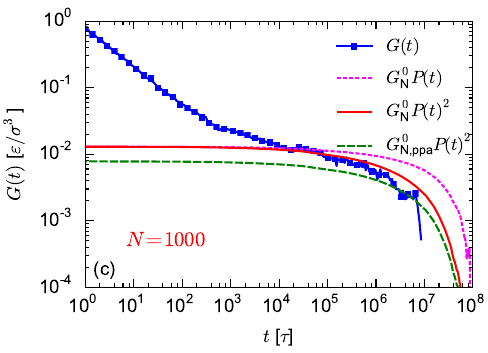}
     
    \caption{Simulation data for $G(t)$ vs. $t$ (blue)
    compared to $G^0_\text{N} P(t)$ (dashed magenta),
    $G^0_\text{N} P(t)^2$ (solid red) and
    $G^0_\text{N,ppa} P(t)^2$ (dashed green) 
    for chain lengths (a) $N = 200$, (b) $400$, and (c)
    $1000$. 
   }
    \label{Fig:Gt-pt} 
\end{figure}

In the absence of CR,  $G(t)$ at $t \gg \tau_\text{e}$ is proportional to the surviving tube fraction, $\mu(t)$~\cite{doi1988theory,mcleish2002tube}. This means that, 
stress is sustained by the part of the chain still inside the tube and, in the absence of CR, is only relaxed by escaping from the original tube.
In this case, we have $G(t)/G^0_{\text{N}} = \mu(t) = P(t)$.
However, for the description of $G(t)$, CR effects cannot be ignored. In the Rouse-tube model, CR leads to a Rouse-like motion of the tube and even small-scale Rouse motions of the tube contribute to the relaxation of stress~\cite{mcleish2002tube,rubinstein2003polymer}. 
Combined
rheological and dielectric experiments on monodisperse linear polymer
melts have suggested the relation~\cite{matsumiya2000comparison,
watanabe2001dielectric}
$G(t)/G^0_{\text{N}} \approx P(t)^{1 + \alpha}$ with $\alpha = 1$, instead of $\alpha = 0$ as in the case of ignoring CR.
This relation expresses the ``double
reptation''~\cite{des1988double,des1990relaxation} and ``dynamic tube
dilation''~\cite{matsumiya2000comparison, watanabe2001dielectric}
approximations, which have also been
explained based on the Rouse-tube model of CR~\cite{mcleish2002tube} (a more detailed discussion of the prediction of the Rouse-tube model is also provided below, in the context of \cref{Eq:Gt-LM}).
The proportionality of $G(t)$ and $P(t)^2$ has also been observed via the simulation of a slip-link model~\cite{masubuchi2001brownian}.
\cref{Fig:Gt-pt} compares $G(t)$ with $P(t)$ and $P(t)^2$ for $N =
200$, $400$, and $1000$.  In all cases, $G^0_{\text{N}}P(t)^2$
(with $G^0_\text{N}=0.013
\varepsilon/\sigma^3$; solid red lines) describes the late-time behavior of $G(t)$
reasonably well, and $G^0_{\text{N}}P(t)$ (dashed magenta lines), which ignores CR, decays
significantly more slowly 
than $G(t)$. The figure also shows that the late time behavior of
$G(t)$ is not captured well if one uses $G^0_\text{N,ppa}$ as prefactor
instead of $G^0_\text{N}$.

We continue the discussion with a comparison of the measured $G(t)$
data with the Likhtman-McLeish model which accounts for the reptation,
CLF, and CR processes, and also the initial short-time Rouse
dynamics~\cite{likhtman2002quantitative}:
\begin{equation}
 \begin{aligned}
         G(t) &= G_\text{e}[ \frac{1}{Z} \sum_{p=Z}^{N} 
           \exp(-\frac{2p^2 t}{\tau_\text{R}})\\
         &+ \frac{1}{5Z} \sum_{p=1}^{Z-1} 
           \exp(-\frac{p^2 t}{\tau_\text{R}})  
           + \frac{4}{5} \mu^\text{LM}(t) \: R(t) ].
    \label{Eq:Gt-LM}  
 \end{aligned}
\end{equation}
Here $G_\text{e} = G(\tau_\text{e}) = \rho k_\text{B} T/N_\text{e}$
and $\mu^\text{LM}(t)$ is the surviving tube fraction which is
calculated using \cref{Eq:LM}. The first term of \cref{Eq:Gt-LM}
describes the initial Rouse motion, the second term is the
contribution of the relaxation of longitudinal modes (a CLF effect
causing redistribution of segments along the tube), and the third term
accounts for
the contribution of escaping from the tube via $\mu^\text{LM}(t)$ and
for the effect of CR via the function $R(t)$.
This latter function was originally introduced by Rubinstein and
Colby~\cite{rubinstein1988self} within a Rouse-tube model of CR and
must be determined self-consistently, assuming that the tube can be
modeled as an effective Rouse chain with random bead mobilities 
whose distribution depends on $\mu(t)$ in a self-consistent manner. 
Likhtman and McLeish introduced an additional fitting parameter, $c_\nu$, 
which controls the decay rate of $R(t)$ and therefore the strength of CR ($c_\nu=1$ corresponds to the
original Rubinstein and
Colby model, $c_\nu=0$ to the total neglect of CR). Here we
use $c_\nu=1$ and determine $R(t)$ using the algorithm proposed by
Rubinstein and Colby~\cite{rubinstein1988self}. 
For $t < \tau_\text{R}$, the functional form of $R(t)$ is
similar to $\mu^\text{LM}(t)$
(\cref{Eq:LM}), however, with a different perfecter: $R(t)= 1 -
(1.8/Z)(t/\tau_\text{e})^{1/4}$. At $t > \tau_\text{R}$, $R(t)$
decays more slowly than $\mu^\text{LM}(t)$ and exhibits a tail. Figure S4 of SI shows $\mu^\text{LM}(t)$ and $R(t)$ together (for $Z = 20$).

\begin{figure}[!htb]
    \centering
    
        \includegraphics[width=0.45\textwidth]{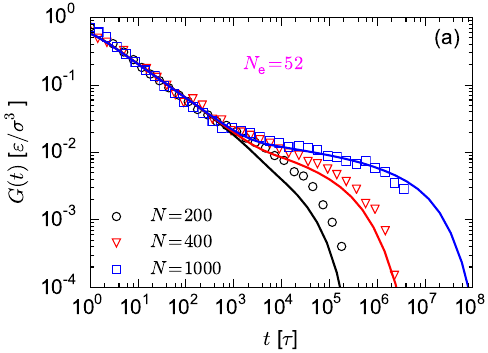}
    
        \includegraphics[width=0.45\textwidth]{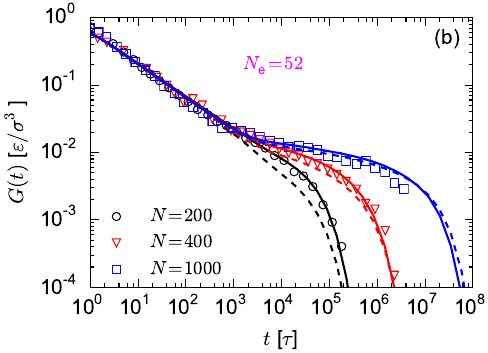}

    \caption{Simulation data for $G(t)$ (symbols) for chain
    lengths
        $N = 200$, $N =400$, and $N = 1000$ compared to the
        predictions of the Likhtman-McLeish model with $N_\text{e} = 52$.
    Solid lines in panel (a) show the result of using the original
    $R(t)$ function of the model with $c_\nu = 1$.  Panel (b) shows
    the results of using the double reptation approximation for CR. In
    this panel, the dashed lines were calculated assuming 
    $R(t)=\mu^\text{LM}(t)$ with $\mu^\text{LM}(t)$ taken from
    \protect\cref{Eq:LM}. The solid lines were calculated by replacing
    $\mu(t)R(t)$ in \cref{Eq:Gt-LM} with $P(t)^2$. 
    }
    \label{Fig:Gt-LM} 
\end{figure}

After setting $c_\nu=1$, \cref{Eq:Gt-LM} has three input
parameters: $N_\text{e}$, $\tau_\text{e}$, and $\tau_\text{R}$.
$\tau_\text{e}$ and $\tau_\text{R}$ can be calculated
using $\tau_\text{e} = \tau_0 N_\text{e}^2$ and  $\tau_\text{R} =
\tau_0 N^2$ and if $\tau_0$ is known
(for the current model $\tau_0 = 1.47 \tau$),
$N_\text{e}$ is the only free parameter. 
Motivated by the discussion above about the value of the plateau modulus we set $N_\text{e}=52$. 
Therefore, there is no adjustable parameter. 
\cref{Fig:Gt-LM}a shows the prediction of the Likhtman-McLeish
model, \cref{Eq:Gt-LM}, for $G(t)$ at chain lengths $N=200$, $N=400$,
and $N=1000$. Following the trend already observed in $P(t)$
(\cref{Fig:LM}), the model does not capture the behavior of $G(t)$
very well for $N=200$, but the quality of the prediction improves with
increasing $N$.

To rationalize this behavior, a short discussion about the general behavior of the model is worth mentioning. 
At late times,
\cref{Eq:Gt-LM} simplifies to  $G(t) = G_0^\text{N} \mu(t) R(t)$.
However, recall that 
at long times, approximately, $G(t) = G_0^\text{N} P(t)^2$ (see \cref{Fig:Gt-pt}).
Therefore, the model produces a good prediction of $G(t)$ when $\mu^\text{LM}(t) R(t)$  is almost equal to $P(t)^2$.
Furthermore, in the range of chain length studied here, $R(t)$ is not far from $\mu^\text{LM}(t)$ (see Figure S4), and for the current discussion about the general behavior of the model, it can be roughly estimated with $\mu^\text{LM} (t)$; therefore, the prediction of the model is roughly $G(t) \approx G_0^\text{N} \mu^\text{LM}(t)^2$. Hence, the model makes a good prediction
 when $\mu^\text{LM}(t)$ is close to $P(t)$. Consistent with this explanation, the prediction of the model for the $G(t)$ of $N = 200$ is not good, because the $\mu^\text{LM}(t)$  does not properly describe the $P(t)$ of $N = 200$ (see the discussion of \cref{Fig:LM}). 
As mentioned above, in the Likhtman-McLeish model, $R(t)$ contains a fitting parameter, $c_\nu$, which is set to be equal to $1$ in the current work. By assuming $c_\nu$ to be chain length-dependent and using  $c_\nu \ll 1$ for $N = 200$ and $N = 400$, it is possible to change the decay rate of $R(t)$ and, consequently, to improve the quality of fitting significantly (Figure S5 shows the output of the model with $c_\nu =0.1$). However, such 
a modification only enables the model to fit the simulation results and does not improve its predictive power.

The results presented in  \cref{Fig:Gt-pt} (comparison between $G(t)$ and $P(t)^2$) suggest taking into account the CR mechanism via the double reptation approximation. 
In \cref{Fig:Gt-LM}b, the dashed lines show the results of assuming $R(t) = \mu^\text{LM}(t)$. 
As mentioned above, $R(t)$ calculated based on the Rouse-tube model of CR is close to $\mu^\text{LM}(t)$, but it is not identical to $\mu^\text{LM}(t)$; particularly, $R(t)$ has a long-time tail (see Figure S4).
The calculated $G(t)$ curves based on $R(t) = \mu^\text{LM}(t)$ assumption are also close to those calculated based on the original $R(t)$ of the model, and the current simulation results are not decisive for comparing the accuracy of these two methods of treating the CR mechanism.
The solid lines in \cref{Fig:Gt-LM}b are also  the results of double relation approximation, however, 
both $\mu(t)$ and $R(t)$ are replaced with $P(t)$ (\ie, $R(t)\mu(t)$ in \cref{Eq:Gt-LM} is replaced by $P(t)^2$ and    \cref{Eq:LM} has not been used for the calculations).
Based on this assumption, good estimates of $G(t)$ can be provided for all $N$,  from the short-time Rouse regime up to the terminal time. This is consistent with the results shown in \cref{Fig:Gt-pt}. 
To show the effect of $N_\text{e}$, in Figure S6 we also present the results
of the replacement $\mu(t) R(t) \to P(t)^2$, but using $N_\text{e}
= N_\text{e,ppa} = 87$. Consistent with the trend observed in \cref{Fig:Gt-pt},
the resulting curves deviate significantly from the
simulation data.

The above approximation (replacement of $\mu(t) R(t)$ with $P(t)^2$) is similar in spirit to an approach proposed by
Hou \etal~\cite{hou2010stress}, who replaced $\mu(t) R(t)$
by $\mu(t)^2$ in \cref{Eq:Gt-LM} according to the double reptation
approximation and estimated $\mu(t)$ by
simulations from the autocorrelation function of the end-to-end vectors of the primitive chains (chain conformation averaged
over a period $\tau_\text{e}$~\cite{read2008entangled}). Combining this method with the use
of $N_\text{e,ppa}$ for the entanglement length,
they also obtained excellent agreement between the predicted and
measured values of $G(t)$ for short and mildly entangled chains. 
However, the results presented in the current work and also the results of Hou \etal~\cite{hou2010stress}
for long chains, suggest that $G_\text{N,ppa}^0 = 4/5 \rho k_\text{B} T/N_\text{e,ppa}$ underestimates the plateau modulus.
Furthermore, by using $N_\text{e} = 52$ we can simplify the overall picture and reveal a simple relation between $G(t)$ and $P(t)^2$ even for mildly entangled chains (instead of a relation between $G(t)$ and end-to-end motion of the primitive chains).

\begin{figure}[!htb]
    \centering
    
        \includegraphics[width=0.45\textwidth]{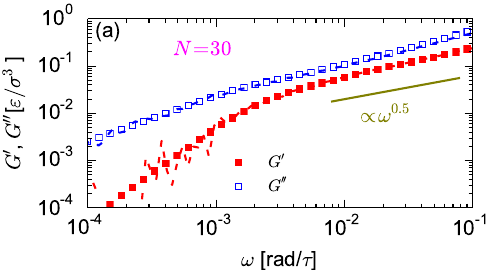}

        \includegraphics[width=0.45\textwidth]{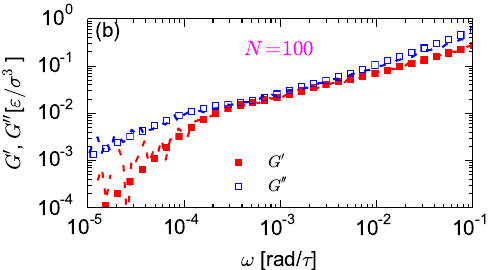}

        \includegraphics[width=0.45\textwidth]{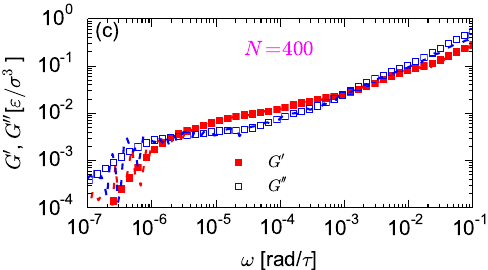}

    \caption{Storage and loss moduli, $G^{\prime}(\omega)$ and
    $G^{\prime\prime}(\omega)$, for chains of length (a) $N =30$,
    (b) $N = 100$, and (c) $N = 400$. The symbols show the moduli
    calculated by fitting a sum of exponential functions to
    $G(t)$. The dashed lines show the moduli calculated via the direct
    Fourier transform of $G(t)$.}

    \label{Fig:stor-loss} 
\end{figure}

From $G(t)$, we can also calculate the storage and loss moduli of
oscillatory shear, $G^{\prime}(\omega)$ and
$G^{\prime\prime}(\omega)$, by Fourier transform.
By fitting $G(t)$ with a series of exponential functions (Maxwell modes), the transformation can be performed analytically~\cite{tassieri2018rheo,behbahani2021dynamics}:
if $G(t) = \sum_i g_i \exp(-t/\lambda_i)$ then 
$G^{\prime}(\omega) = \sum_i g_i  \lambda_i^2 \omega^2/(1+ \lambda_i^2 \omega^2)$ and $G^{\prime\prime}(\omega) = \sum_i g_i  \lambda_i \omega/(1+\lambda_i^2 \omega^2 )$.

\cref{Fig:stor-loss} (symbols) show the $G^{\prime}(\omega)$ and $G^{\prime\prime}(\omega)$ calculated through fitting exponential functions, for $N = 30$, $100$, and $400$, for frequencies smaller than those of the segmental dynamics and bond vibrations.
For the
fitting, we ignored the short-time oscillations of $G(t)$ due to bond
vibrations and fitted the Maxwell modes in the range $t > 0.01 \tau$.
For comparison, the figure also shows the results obtained by a direct Fourier transform
of the bare data using the Reptate package~\cite{tassieri2018rheo,boudara2020reptate} (dashed lines),
which agree with the exponential fit data, but are
much more noisy in the low-frequency limit.
The $G^{\prime}(\omega)$ and $G^{\prime\prime}\omega)$ curves
feature the signatures of the transition from unentangled to
entangled dynamics.
For moderately entangled chains with length $N=400$, two crossing
points are observed between $G^{\prime}(\omega)$ and
$G^{\prime\prime}(\omega)$; also, at frequencies higher than the high-frequency cross-over point   
$G^{\prime\prime}(\omega) > G^{\prime}(\omega)$.
For $N = 100$, the moduli curves are almost tangent, 
and for $N = 30$, the intersection of $G^{\prime}(\omega)$ and $G^{\prime\prime}\omega)$  is not observed.   
The high-frequency intersection point between
$G^{\prime}(\omega)$ and $G^{\prime\prime}(\omega)$ provides an
estimate of $\tau_\text{e} \approx
1/\omega_\text{cross}$~\cite{tassieri2018rheo}. For $N = 400$,
we obtain
$1/\omega_\text{cross} \approx 800 \tau$, which is smaller than 
 other estimates of $\tau_\text{e}$, including the one based on
$G(\tau_\text{e}) = \rho k_\text{B} T/N_\text{e}$ discussed above.

We note that Likhtman \etal~\cite{likhtman2007linear} performed a similar analysis, also using the Kremer-Grest model, and did not
observe a high-frequency intersection between $G^{\prime}(\omega)$ and
$G^{\prime\prime}(\omega)$ for entangled chains. This discrepancy
is probably due to differences in the fitting procedures.  According
to our experience with the Kremer-Grest model and also other
atomistic and coarse-grained polymer
models~\cite{behbahani2021dynamics}, 
the high-frequency cross-over point is  observed when time scales smaller than the timescale of segmental dynamics are included
in the fitting range (that is, the drop of $G(t)$ due to segmental
dynamics must be included).

\begin{figure}[t]
    \centering
    
        \includegraphics[width=0.45\textwidth]{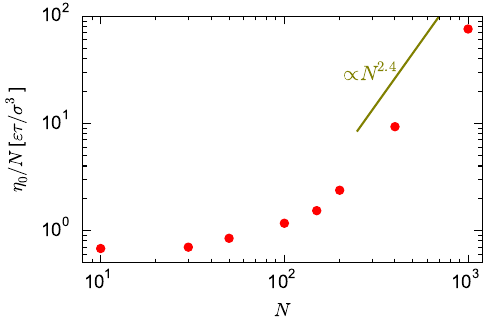}
    \caption{ Zero-shear viscosity divided by chain length, $\eta_0/N$, versus $N$.}
    \label{Fig:eta} 
\end{figure}

We conclude the discussion of the viscoelastic properties by
analyzing the zero-shear viscosity, which can be calculated from
the integral of $G(t)$, $\eta_0 = \int_0^\infty G(t)\mathrm{d}t$. The
integral was calculated analytically after fitting  $G(t)$ with a
series of exponential functions as described above. For the
longest chains with $N=1000$, the data for $G(t)$ do not have good
statistics at late times; for this case, the long time behavior of
$G(t)$ was estimated by $G^0_\text{N} P(t)^2$ (see
\cref{Fig:Gt-pt}).  \cref{Fig:eta} shows $\eta_0/N$ versus $N$.
One can clearly see the transition from Rouse behavior ($\eta_0
\sim N$) to the entangled regime. However, in the range of studied
chain lengths, a small deviation from  the expected scaling exponent $\approx 3.4$, which
would be consistent with the exponent observed for
$\tau_\text{d}$ (\cref{Fig:pt-tau}),  is observed.  For
very large $N$, the contribution of the short-time Rouse dynamics is
negligible, and $\eta_0$ is dominated by the entangled dynamics with a
characteristic time of $\tau_\text{d}$. However, for moderately
entangled chains, the contribution of the Rouse regime, which has a
$N$-independent characteristic time of $\tau_\text{e}$, can not be
ignored. This contribution leads to a deviation of the scaling
exponent of $\eta_0$ from the scaling exponent of $\tau_\text{d}$ for
moderately entangled chains.
 
\section{Summary}

We have presented a detailed analysis of the relaxation dynamics
of linear polymer melts using molecular dynamics simulations.
Simulations were performed using the standard Kremer-Grest bead-spring
model of chains with different lengths, ranging from $N = 10$ up
to $N = 1000$. This latter $N$ corresponds to $Z \approx 20$
entanglements per chain based on $N_\text{e} = 52$ as
estimated from the plateau modulus.  We focused on the analysis of
mean-squared displacements ($g_1(t)$ and $g_3(t)$), the
 autocorrelation function of the end-to-end vector
($P(t)$), the single-chain dynamic structure factor ($S(q,t)$),
and linear viscoelastic properties, particularly the shear
stress-relaxation modulus ($G(t)$). 
Based on the results of the
single-chain dynamic structure factor, we also computed
wave-vector dependent relaxation rates $\tau(q)$ and non-local
mobility functions, $\Lambda(q)$, that can be used in dynamical
density functional theory calculations of the dynamics of structure
evolution in heterogeneous polymer melts. 

Special attention was given to detecting the signs of different
relaxation mechanisms of entangled melts, namely CLF (contour length
fluctuation), CR (constraint release), and reptation, and to the
comparison of the simulation data with analytical theories based on
the tube model and on the concept of the ``surviving tube fraction''
$\mu(t)$.   Our main results can be summarized as follows:
\begin{itemize}
\item With increasing chain length, gradual development of scaling regimes $(1-P(t))\sim t^{1/4}$
and $\chi^{\prime\prime}(\omega) \sim \omega^{-1/4}$ (where $\chi^{\prime\prime}(\omega)$ is dynamic susceptibility and shows end-to-end dynamics on the frequency domain) is observed. 
These regimes can be explained based on CLF which is
responsible for the appearance of a scaling regime  $(1-\mu(t)) \sim t^{1/4}$ 
at intermediate times. 
Neglecting the CR effect, we have $P(t)=\mu(t)$, and the above scaling regimes are consistent with the expectation based on CLF.

\item A general agreement between the time dependence of $g_1(t)$ and the time dependence of $(1-P(t))$ is observed.
Particularly, because of the restricted Rouse motion of the chain in their tubes, both quantities scale with $t^{1/4}$ in the interval of $\tau_\text{e} < t < \tau_\text{R}$.

\item We observed signatures of the CR process in the relaxation of $S(q,t)$ and $G(t)$.
In the absence of CR, the tube theory predicts 
$S(q,t) \propto G(t) \propto P(t) =  \mu(t)$ at $t \gg \tau_\text{e}$ and $2\pi/R_\text{ee} \ll q \ll 2\pi/a$. 
For both $S(q,t)$ and $G(t)$, proportionality with $P(t)$ was not observed; both quantities decay significantly faster than $P(t)$.
Instead, we find $G(t) \propto P(t)^2$, 
consistent with the dynamic tube dilation or double
reptation approximations.

\item One can estimate the plateau modulus and its corresponding entanglement length from the $G(t)$ data
using the following self-consistency condition: for a given $N_\text{e}$ value one can calculate a $\tau_\text{e}$ from the  $G(t)$ curves using $G(\tau_\text{e}) = \rho k_\text{B} T/N_\text{e}$; this $\tau_\text{e}$ value is expected to be consistent with the Rouse time of $N_\text{e}$ monomers, \ie, $\tau_\text{e} = \tau_0 N_\text{e}^2$. 
In the case of the fully flexible Kremer-Grest polymer melts studied here, we found $N_\text{e} = 52$, which was previously calculated using an equivalent single-chain slip-link model~\cite{likhtman2007linear}, and is much smaller than the value $N_\text{e,ppa} = 87$ obtained by 
the primitive path analysis, satisfies the above condition. 
\item The measured $P(t)$ and $S(q,t)$ data were compared with the predictions of the pure reptation 
and the predictions based on CLF and reptation. 
In both cases, pure reptation does not properly describe the simulation results.
In the case of $P(t)$, the inclusion of CLF
significantly improves the quality of theoretical predictions. 
However, even the CLF model did not provide
a satisfactory description of the $S(q, t)$ data. 
This signals the importance of CR for the relaxation of $S(q,t)$, at least in the range of chain lengths studied here.

\item The simulation results for $G(t)$ were compared with the predictions of the
Likhtman-McLeish~\cite{likhtman2002quantitative} model which takes into account reptation, CLF, and CR mechanisms.
The model does not describe the simulation results for mildly entangled chains, however, the quality of the prediction improves with increasing chain length.
At late times, the model assumes a simple functions form: $G(t) = G^0_\text{N} \mu(t) R(t)$, where $R(t)$ accounts for the contribution of CR. 
Motivated by the relation $G(t) = G^0_\text{N} P(t)^2$ (mentioned above), we replaced 
$\mu(t) R(t)$ term in the model with $P(t)^2$ and obtained excellent agreement between the resulting curves and the simulation data for $G(t)$. However, in this case, only the short-time behavior of $G(t)$ is 
constructed based on theoretical expressions. 

\end{itemize}

\subsection*{Supporting Information}

The Supporting Information contains:
(1) Mean-squared displacements of the centers-of-mass of the chains. (2) Description of $P(t)$ based on $N_\text{e,ppa}$. (3) Theoretical predictions for the $S(q,t)$ of $N = 400$. (4) Comparing $R(t)$ and $\mu(t)$ of the Likhtman-McLeish model. (5) The effect of $c_\nu$ on the output of the Likhtman-McLeish model. (6) Description of $G(t)$ based on $N_\text{e,ppa}$.

\subsection*{Acknowledgements}
 This research was supported by the German Science Foundation
 (DFG) via SFB TRR 146 (Grant number 233630050, project C1). 
 The simulations were carried out on the high
 performance computing center MOGON at JGU Mainz.

\end{document}


\title{Supporting Information: 
Relaxation Dynamics of Entangled Linear Polymer Melts via Molecular Dynamics Simulations}

\author{Alireza F. Behbahani}
\email{aforooza@uni-mainz.de}
\affiliation{Institut f\"{u}r Physik, Johannes Gutenberg-Universit\"{a}t Mainz, Staudingerweg 7, D-55099 Mainz, Germany}

\author{Friederike Schmid}
\email{friederike.schmid@uni-mainz.de}
\affiliation{Institut f\"{u}r Physik, Johannes Gutenberg-Universit\"{a}t Mainz, Staudingerweg 7, D-55099 Mainz, Germany}

\maketitle

\textbf{Mean-squared displacements of the centers-of-mass of the chains, $g_3(t)$}

\begin{figure}[!htb]
    \centering
    \begin{subfigure}{0.45\textwidth} 
        \includegraphics[width=\textwidth]{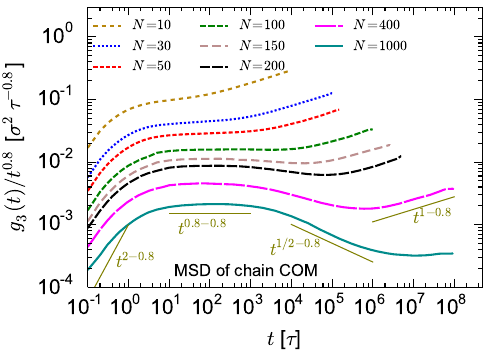}
    \end{subfigure}
\caption{Mean-squared displacement of the centers-of-mass of the chains divided by $t^{0.8}$.
  }
  \label{Fig:sm-g3-normal}
\end{figure}

Figure~\ref{Fig:sm-g3-normal} shows $g_3(t)$ divided by $t^{0.8}$ for different chain lengths. 
After the very short time ballistic regime, $g_3(t)$ of all chains scale almost with $t^{0.8}$. This is reflected in the appearance of a short-time plateau regime in the $g_3(t)/t^{0.8}$ curve.
After this regime, $g_3(t)/t^{0.8}$ of entangled chains ($N > 100$), exhibit a negative slope. With increasing $N$, the negative slope tends to the prediction of the tube model ($g_3(t)\sim t^{0.5}$ or $g_3(t)/t^{0.8}\sim t^{-0.3}$).

\bigskip
\textbf{Description of $P(t)$ based on $N_\text{e,ppa}$}

\begin{figure}[htb!]
    \centering
    \begin{subfigure}{0.45\textwidth} 
        \includegraphics[width=\textwidth]{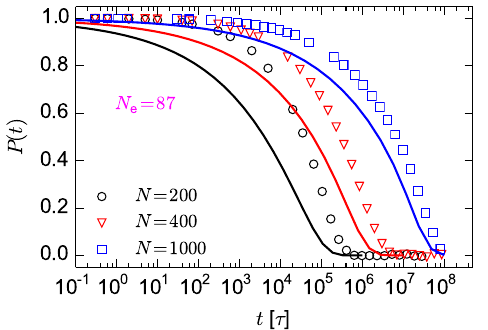}
    \end{subfigure}
    \caption{$P(t)$ for $N = 200$, $400$, and $1000$ (symbols) together with the $\mu(t)$ of Likhtman-McLeish model (solid lines) with $N_\text{e,ppa} = 87$.}
    \label{Fig:sm-pt} 
\end{figure}

The symbols
in Figure~\ref{Fig:sm-pt}
show the the autocorrelation function of the end-to-end vector, $P(t)$, for $N =200$, $400$, and $1000$.
The solid lines show the  result of the 
Likhtman-McLeish model~\cite{likhtman2002quantitative} ($\mu^\text{LM}(t)$, Equation 9 of the main text) using $N_\text{e} = N_\text{e,ppa} = 87$.
As Figure~\ref{Fig:sm-pt} shows, using $N_\text{e,ppa}$ does not lead to a good description of the measured $P(t)$ data, as compared to the results of using $N_\text{e} = 52$ shown in Figure 5 of the main text.

\clearpage
\bigskip
\textbf{Theoretical predictions for the $S(q,t)$ of $N = 400$}

\begin{figure}[htb!]
    \centering
    \begin{subfigure}{0.45\textwidth} 
        \includegraphics[width=\textwidth]{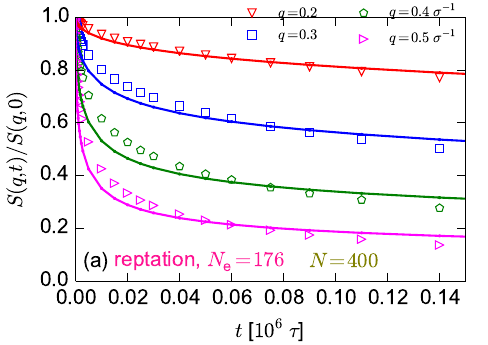}
    \end{subfigure}
    \begin{subfigure}{0.45\textwidth} 
        \includegraphics[width=\textwidth]{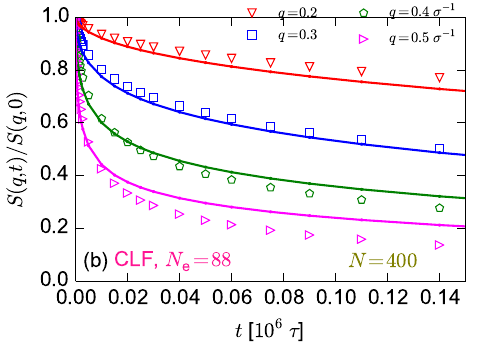}
    \end{subfigure}

    \begin{subfigure}{0.45\textwidth} 
        \includegraphics[width=\textwidth]{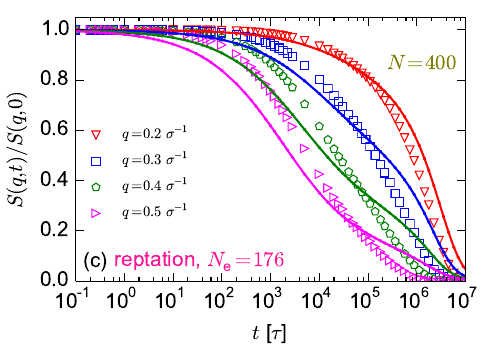}
    \end{subfigure}
    \begin{subfigure}{0.45\textwidth} 
        \includegraphics[width=\textwidth]{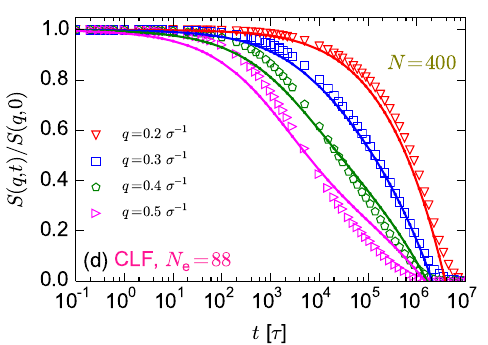}
    \end{subfigure}

\caption{The symbols show the simulation results for  $S(q,t)/S(q,0)$ of $N = 400$ for different values of $q$. Solid lines in panels (a) and (c) show the theoretical predictions for $S(q,t)/S(q,0)$ based on the pure reputation model with $N_\text{e} = 176$, in linear and semi-logarithmic scales. Panels (b) and (d) show the theoretical predictions based on CLF  with $N_\text{e} = 88$.
}
    \label{Fig:sm-sq} 
\end{figure}

Figure~\ref{Fig:sm-sq} shows the the measured $S(q, t)$ for $N = 400$, together with the predictions based on the pure reptation and CLF (contour length fluctuation) processes.
Similar to the trend observed for $N = 1000$ (Figure 11 of the main text), the CLF model describes the $S(q,t)$ data better and with a smaller apparent entanglement length compared to the pure reptation model.

\clearpage
\textbf{Comparing $R(t)$ and $\mu(t)$ of the Likhtman-McLeish model}

\begin{figure}[!htb]
    \centering
    \begin{subfigure}{0.45\textwidth} 
        \includegraphics[width=\textwidth]{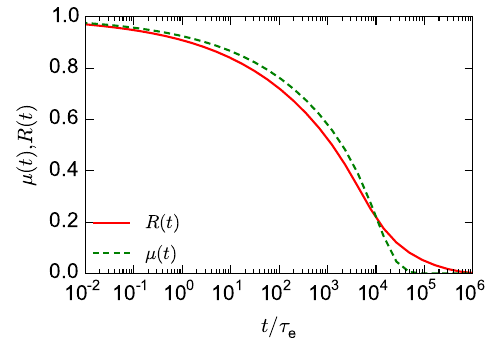}
    \end{subfigure}
\caption{$R(t)$ and $\mu(t)$ of the Likhtman-McLeish model for $Z = N/N_\text{e} = 20$.
}
    \label{Fig:sm-rt} 
\end{figure}

Figure~\ref{Fig:sm-rt} shows the $\mu(t)$ and $R(t)$ functions of the Likhtman-McLeish model~\cite{likhtman2002quantitative} 
for $Z = N/N_\text{e} = 20$.
$R(t)$ is calculated based on an algorithm proposed by Rubinstein and
Colby~\cite{rubinstein1988self} based on a Rouse-tube model of CR ($c_\nu = 1$). 
At $t> \tau_\text{R}$, $R(t)$ decays slower than $\mu(t)$ and exhibits a tail.

\bigskip
\textbf{The effect of $c_\nu$ on the output of the Likhtman-McLeish model}

\begin{figure}[!htb]
    \centering
    \begin{subfigure}{0.45\textwidth} 
        \includegraphics[width=\textwidth]{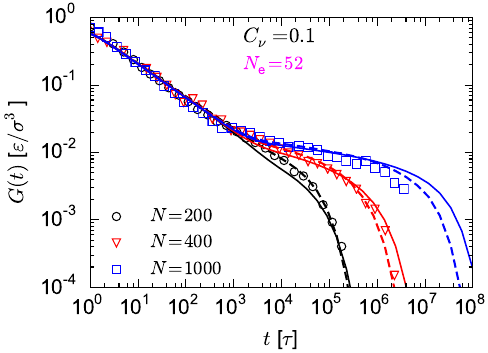}
    \end{subfigure}
\caption{Symbols: the simulation results for the $G(t)$ of $N = 200$, $N = 400$, and $N = 1000$. Solid line: the prediction of the Likhtman-McLeish model with $N_\text{e} = 52$ and $c_\nu = 0.1$. Dashed lines: the prediction based on replacing $\mu(t)R(t)$ by $P(t)^2$ in the Likhtman-McLeish model.}
\label{Fig:sm-cv}
\end{figure}

Figure~\ref{Fig:sm-cv} shows the prediction of the Likhtman-McLeish model~\cite{likhtman2002quantitative} with  $c_\nu = 0.1$ together with the prediction obtained by replacing 
the $\mu(t)R(t)$ term in the model by $P(t)^2$.
At long times the Likhtman-McLeish model has a simple functional form: $G(t) = G^0_\text{N} \mu(t)R(t)$. 
One can adjust the decay rate of $R(t)$ by changing an adjustable parameter, $c_\nu$, and as a result, change the resulting $G(t)$ function.

For $N = 1000$, $c_\nu = 1$ seems to produce a reasonable prediction (see Figure 15 of the main manuscript). However, with $c_\nu = 0.1$, the predicted $G(t)$ for $N = 1000$
is slower than expected (that is $G(t)$ is slower than $G_\text{N}^0 P(t)^2$; see Figure 14 of the main manuscript).
For $N = 400$ and $N = 200$, the predictions based on $c_\nu = 0.1$ are better than those with $c_\nu = 1$, shown in Figure 15 of the main manuscript.
For these chain lengths, $\mu(t)$ of the Likhtman-McLeish model is faster than the corresponding $P(t)$ curves (Figure 5 of the manuscript); to compensate for the faster decay of $\mu(t)$, we need $c_\nu \ll 1$ to reduce the decay rate of $R(t)$ and make $\mu(t) R(t)$ comparable to $P(t)^2$ and therefore produce good predictions for $G(t)$.

Overall, by assuming $c_\nu$ to be chain length-dependent it is possible to change the behavior of $R(t)$ and, consequently, to improve the quality of fitting. However, such a modification only enables the model to fit the simulation results and does not improve its predictive power.

\bigskip
\textbf{Description of $G(t)$ based on $N_\text{e,ppa}$}

\begin{figure}[!htb]
    \centering
    \begin{subfigure}{0.45\textwidth} 
        \includegraphics[width=\textwidth]{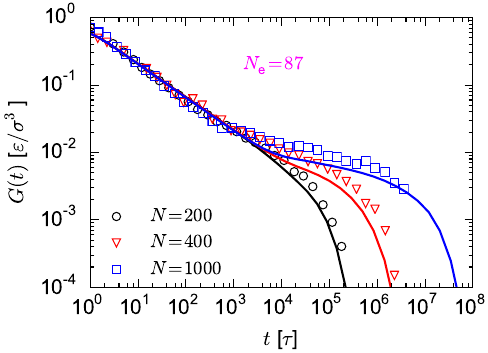}
    \end{subfigure}
\caption{The simulation results (symbols) for the $G(t)$ of $N = 200$, $N =400$, and $N = 1000$ together with theoretical estimations (solid lines) by assuming that the plateau modulus can be calculated based on $N_\text{e,ppa} = 87$.
The solid lines have been calculated based on the Likhtman-McLeish model and double reptation approximation for CR, $R(t) = \mu(t) = P(t)$, where $P(t)$ is the autocorrelation function of the end-to-end vector. 
}
    \label{Fig:sm-gt} 
\end{figure}

Solid lines in Figure~\ref{Fig:sm-gt} show the 
the $G(t)$ curves calculated for $N = 200$, $N = 400$, and $N = 1000$ by assuming that the entanglement length is equal to $N_\text{e,ppa} = 87$.
For the calculation of the solid lines, the $\mu(t) R(t)$ term in the Likhtman-McLeish model~\cite{likhtman2002quantitative} (Equation 18 of the main text) was replaced with $P(t)^2$ (see the discussion of Figure 15 of the main text). 
As can be seen from Figure~\ref{Fig:sm-gt}, the  $G(t)$ curves estimated based on $N_\text{e,ppa} = 87$ do not agree with the measured $G(t)$ data (shown with symbols).